\documentclass[11pt,a4paper]{article}
\pdfoutput=1
\usepackage{jheppub}
\usepackage{amsmath,amssymb,amsfonts}
\usepackage{graphicx,graphics}
\usepackage{graphics}
\usepackage{url}
\usepackage{slashed}
\usepackage{color}
\usepackage{mathtools}
\usepackage[utf8]{inputenc}

\title{B-meson hadroproduction in the SACOT-$m_{\rm T}$ scheme}

\affiliation[a]{University of Jyvaskyla, Department of Physics, P.O. Box 35, FI-40014 University of Jyvaskyla, Finland}
\affiliation[b]{Helsinki Institute of Physics, P.O. Box 64, FI-00014 University of Helsinki, Finland}

\emailAdd{ilkka.m.helenius@jyu.fi}
\emailAdd{hannu.paukkunen@jyu.fi}

\abstract{
We apply the SACOT-$m_{\rm T}$ general-mass variable flavour number scheme (GM-VFNS) to the inclusive B-meson production in hadronic collisions at next-to-leading order in perturbative Quantum Chromodynamics. In the GM-VFNS approach one matches the fixed-order heavy-quark production cross sections, accurate at low transverse momentum ($p_{\rm T}$), with the zero-mass cross sections, accurate at high $p_{\rm T}$. The physics idea of the SACOT-$m_{\rm T}$ scheme is to do this by accounting for the finite momentum transfer required to create a heavy quark-antiquark pair throughout the calculation. We compare our results with the latest LHC data from proton-proton and proton-lead collisions finding a very good agreement within the estimated theoretical uncertainties. We discuss also scheme-related differences and their impact on the scale uncertainties.

}


\begin{document}

\author{Ilkka Helenius$^{a,b}$ and}
\author{Hannu Paukkunen$^{a,b}$}

\maketitle

\section{Introduction}
\label{sec:intro}

The production of mesons containing a bottom quark -- collectively called B mesons -- in hadronic collisions provides a useful way to study various aspects of Quantum Chromodynamics (QCD). On one hand, thanks to the large bottom quark mass $m_{\rm b} \approx 4...5\,{\rm GeV}$, the perturbative expansion in powers of the strong QCD coupling $\alpha_s$ can be expected to converge relatively well and thereby provide an accurate description of the production mechanism \cite{Combridge:1978kx,Beenakker:1988bq,Nason:1989zy,Czakon:2013goa,Catani:2020kkl}. In comparison to charm-quark production where the possible non-perturbative intrinsic charm-quark content of the nucleons \cite{Brodsky:1980pb,Ball:2022qks,Guzzi:2022rca} can stir the interpretation, the bottom-quark production can be seen to be a cleaner process to test the perturbative QCD, though an intrinsic bottom-quark component is not excluded \cite{Lyonnet:2015dca, Forte:2019hjc}. The B-meson production is sensitive especially to the gluon content of the colliding hadrons and can thus be used to provide information on their non-perturbative structure, the parton distribution functions (PDFs) \cite{Cacciari:2015fta,Gauld:2015yia,Eskola:2019bgf,Kusina:2020dki}. While the B-meson production is not used as a constraint in the current global fits of proton PDFs \cite{Bailey:2020ooq,NNPDF:2021njg,Hou:2019efy}, it should be mentioned that e.g. in comparison to the jet production \cite{CMS:2016jip,ALICE:2019qyj} -- a commonly used strong gluon constraint -- no external corrections due to multi-parton interactions or hadronization need to be supplied but the entire process can be calculated within the collinear factorization. As a result, the B-meson production could provide a rather clean probe for gluon distributions relying solely on inclusive single-particle production. On the other hand, for observables like the Drell-Yan dilepton or direct W$^\pm$ production the weak decays of heavy-flavoured mesons also produce a significant background of charged leptons whose subtraction requires an accurate theoretical understanding of the heavy-quark production \cite{ALICE:2022cxs}. Analyzing B-meson production in proton-nucleus collisions could provide further constraints for nuclear PDFs and, in the context of heavy-ion collisions, the B-mesons can also be used as a probe of the produced strongly interacting matter \cite{CMS:2017uoy} and the expected mass hierarchies. 

The cross sections for identified B-meson hadroproduction have been measured in several collision systems: proton-antiproton (p-$\overline{\rm p}$) \cite{CDF:2004jtw,CDF:2006ipg} collisions at Fermilab Tevatron, as well as in proton-proton (p-p) \cite{CMS:2011pdu,CMS:2011oft,CMS:2011kew,LHCb:2012sng,LHCb:2013vjr,ATLAS:2013cia,CMS:2016plw,LHCb:2016qpe,LHCb:2017vec}, proton-lead (p-Pb) \cite{CMS:2015sfx,LHCb:2019avm}, and lead-lead (Pb-Pb) \cite{CMS:2017uoy} collisions at the Large Hadron Collider (LHC). In many occasions the B-meson cannot be fully reconstructed but only the spectrum of specific decay particles like charged leptons or J/$\psi$ mesons, are measured. In work presented here, we will concentrate exclusively on the reconstructed B mesons, but plan to return to the decay spectra in future publications. 

We will discuss the B-meson production mainly in the so-called general-mass variable-flavour-number scheme (GM-VFNS) \cite{Thorne:2008xf}. The GM-VFNS provides a framework to complement fixed-order QCD calculations with a resummation of heavy-quark mass-dependent logarithms that arise from collinear splitting of partons to heavy quarks. The fixed-order calculations -- known to leading order (LO) \cite{Combridge:1978kx}, next-to-LO (NLO) \cite{Beenakker:1988bq,Nason:1989zy}, and next-to-NLO (NNLO) \cite{Czakon:2013goa,Catani:2020kkl} in strong coupling $\alpha_s$ -- are based purely on diagrams in which the heavy quarks are explicitly excited from massless partons. The resummed parts account for the possibility that the heavy quarks are produced through higher-order diagrams within the initial- and final-state radiation. Although formally suppressed by extra powers of $\alpha_s$ these contributions arise from collinear configurations which are logarithmically enhanced at large values of transverse momenta ($p_{\rm T}$). The division between the explicit and shower-originating heavy-quark production channels is not unique which induces a scheme and scale dependence on the description. 

Historically, the first variant of GM-VFNS for heavy-flavour hadroproduction was the so-called FONLL (Fixed-Order Next-to-Leading Logarithm) scheme introduced in Ref.~\cite{Cacciari:1998it}. Later on the SACOT (Simplified Aivazis-Collins-Olness-Tung) scheme was presented in Refs.~\cite{Kniehl:2004fy,Kniehl:2005mk} and has been later on applied e.g. in Refs.~\cite{Kniehl:2007erq,Kniehl:2011bk}. In the SACOT scheme, part of the resummed contributions are described by massless partonic coefficient functions which induces an unphysical divergence towards $p_{\rm T} \rightarrow 0$, and one cannot therefore generally extend the calculation down to zero $p_{\rm T}$. In Refs.~\cite{Kniehl:2015fla,Kramer:2018vde,Benzke:2019usl} the authors pointed out that this behaviour can be tamed by suitably tuning the factorization and fragmentation scales. In the FONLL scheme these divergent features are cured by multiplying the zero-mass contributions by a factor ${p_{\rm T}^2}/({p_{\rm T}^2 + c^2m^2})$, where $c = 5$ by default and $m$ is the heavy-quark mass, which serves to evade the unphysical behaviour while still respecting the principles of GM-VFNS. However, neither of the two is a particularly natural way to cure the divergent behaviour and the former also causes unphysical kinks to the $p_{\rm T}$ spectrum of heavy-flavoured mesons. Indeed, the reason why the invariant heavy-quark cross section remains finite even at zero $p_{\rm T}$ is in the mass of the heavy quark which, when properly accounted for, keeps the intermediate particles off-shell -- there is always a finite momentum transfer between the colliding, massless initial-state partons. This is the underlying physics idea of the SACOT-$m_{\rm T}$ scheme which was introduced in Ref.~\cite{Helenius:2018uul}. It is the counterpart of the SACOT-$\chi$ scheme \cite{Guzzi:2011ew} often used in the context of deeply inelastic scattering. Very recently, preliminary documents of the so-called SACOT-MPS (Massive Phase Space) scheme have also appeared \cite{Xie:2019eoe,Xie:2021ycd}, which seems to share partly same ideas as the SACOT-$m_{\rm T}$ scheme applied in this work.

A somewhat different but closely related approach to heavy-flavour hadroproduction is the one in which fixed-order calculations are matched with a parton shower (FO-PS) \cite{Nason:2004rx,Frixione:2007vw,Alioli:2010xd,Mazzitelli:2023znt}. This procedure also performs a similar resummation as done in GM-VFNS though it still, in general, misses part of the resummed contributions that are included in GM-VFNS \cite{Helenius:2018uul}, though it can be used to simulate exclusive final states as well. Also, while it is more natural to use 4-flavour PDFs in the context of FO-PS framework to describe $b\overline{b}$ production (part of the logarithms resummed by the parton shower are included in the evolution of the $b$-quark PDFs) a consistent use of 5-flavour PDFs is a built-in feature of GM-VFNS making it well-suited for general-purpose PDF studies. 

In the present paper our aim is to apply the SACOT-$m_{\rm T}$ scheme \cite{Helenius:2018uul}, originally devised in the context of D-meson (mesons containing a charm quark) production, to the case of B mesons. The differential cross section $\mathrm{d}\sigma / \mathrm{d} p_{\mathrm{T}}$ of both D- and B-mesons show a maximum at low $p_{\rm T}$ but they occur at different values of $p_{\rm T}$. How this is linked with the heavy-quark masses is an intrinsic feature of a given scheme and provides thus a well-defined way to study the reliability of different schemes. We also introduce an improved description of the fragmentation variable which evades some difficulties in the original setup. In what follows we will first introduce the formalism in Section \ref{sec:framework}, and then discuss the numerical results in Sections \ref{sec:results-pp} and \ref{sec:results-pPb} for p-p and p-Pb collisions at the LHC, respectively. In Section~\ref{sec:Conclusion} we summarize the paper discussing our future plans.

\section{The SACOT-$m_{\rm T}$ framework}
\label{sec:framework}

We will now recapitulate our SACOT-$m_{\rm T}$ framework \cite{Helenius:2018uul} for single-inclusive heavy-flavoured meson production in hadronic collisions. The process we study is, 
$$
h_1(P_1) + h_2(P_2) \longrightarrow h_3(P_3) + X \,,
$$
where $h_1$ and $h_2$ denote the colliding hadrons and $h_3$ is the heavy-flavoured meson. The momenta of the hadrons are indicated by $P_i$. We can write the invariant cross section as 
\begin{align}
\frac{\mathrm{d}^3\sigma^{h_1 + h_2 \rightarrow h_3 + X}}{\mathrm{d}^3P_3/P_3^0}
& =
 \sum _{ijk} 
 \int_{z^{\rm min}}^1 \frac{\mathrm{d}z}{z^2} \int_{x_1^{\rm min}}^1 \mathrm{d}x_1 \int _{x_2^{\rm min}}^1 \mathrm{d}x_2 
 f_i^{h_1}(x_1,\mu^2_{\rm fact}) \, f_j^{h_2}(x_2,\mu^2_{\rm fact}) \, D_{k \rightarrow h_3}(z,\mu^2_{\rm frag})
  \nonumber \\[3pt]
& \times  J(\vec p, \vec P) \times
\frac{\mathrm{d}^3\hat{\sigma}^{ij\rightarrow k+X} (\tau_1, \tau_2, \rho , \sqrt{s}, \mu^2_{\rm ren}, \mu^2_{\rm fact}, \mu^2_{\rm frag})}{\mathrm{d}^3p_3/p_3^0} 
\label{eq:master} \\[3pt]
& - {\rm subtractions} \,. \nonumber
\end{align}
Here, $d\hat\sigma^{ij\rightarrow k+X}/d^3p_3$ are the inclusive partonic cross section for producing a parton $k$ carrying a momentum $p_3$ in collisions of partons $i$ and $j$ with momenta $p_1 = x_1P_1$ and $p_2 = x_2P_2$ in our scheme. The fragmentation of the produced parton $k$ into a heavy-flavoured meson is described by the fragmention functions (FFs) $D_{k \rightarrow h_3}(z,\mu^2_{\rm frag})$ which depend on the fragmentation scale $\mu^2_{\rm frag}$. The fluxes of partons from the initial-state hadrons are described by the PDFs $f_i(x,\mu^2_{\rm fact})$ and they depend on the factorization scale $\mu^2_{\rm fact}$. The subtraction terms are required in order avoid the double counting between the same logarithmic terms that appear in partonic cross sections and PDFs/FFs, as will be discussed later on.

The invariants $\tau_1$, $\tau_2$, and $\rho$ are defined by
\begin{align}
\tau_1 \equiv \frac{p_1 \cdot p_3}{p_1 \cdot p_2} = \frac{m_{\rm T}e^{-y}}{x_2 \sqrt{s}} 
\,, \ \ \ \
\tau_2 \equiv \frac{p_2 \cdot p_3}{p_1 \cdot p_2} = \frac{m_{\rm T}e^{y}}{x_1 \sqrt{s}} 
\,, \ \ \ \
\rho \equiv \frac{m^2}{x_1x_2s}  \,, 
\label{eq:tau12}
\end{align}
where $m_{\rm T} = \sqrt{p_{\rm T}^2 + m^2}$ and $y$ denote the transverse mass and rapidity of the parton $k$. Here, $p_{\rm T}$ is the partonic transverse momentum and $m$ is the heavy-quark mass. The integration limits $x_{1,2}^{\rm min}$ are 
\begin{equation}
  x_1^{\rm min} = \frac{m_{\rm T} \, e^{ y}}{\sqrt{s}-m_{\rm T} \, e^{-y}}, \quad
  x_2^{\rm min} = \frac{x_1 m_{\rm T} \, e^{-y}}{x_1\sqrt{s}-m_{\rm T} \, e^{y}} \,. 
\end{equation}

The transverse momentum $P_{\rm T}$ and rapidity $Y$ of the heavy-flavoured meson are related to the corresponding partonic quantities through the definition of the fragmentation variable $z$, for which we now use
\begin{align}
z \equiv \frac{P_3 \cdot \left(P_1 - P_2\right)}{p_3 \cdot \left(P_1 - P_2\right)} \xrightarrow{\rm c.m. \ frame}  \frac{P_{\rm T}}{p_{\rm T}} =  \frac{|\vec P|}{|\vec p|}\,, \label{eq:fragvar}
\end{align}
where we have assumed that the fragmentation is collinear in the center-of-mass (c.m.) frame of the collision. This definition of $z$ is associated with the Jacobian factor in Eq.~(\ref{eq:master}),
\begin{align}
J(\vec p, \vec P) = 
\sqrt{ \frac{ {\vec P_3}^2 + M^2}{{\vec P_3}^2}  
\frac{{\vec p_3}^{\,2}}{{\vec p_3}^{\,2} + m^2}  } \,,
\end{align}
where $M$ is the meson mass, and the integration limit $z^{\rm min}$ is 
\begin{equation}
  z^{\rm min}   = \frac{|\vec P_3|}{\sqrt{s/4-m^2}} \, . \label{eq:ZMmin}
\end{equation}
We note that the definition of the fragmentation variable $z$ in Eq.~(\ref{eq:fragvar}) is a little different than the definition our earlier work \cite{Helenius:2018uul} where we defined the fragmentation variable as $z' \equiv {P_3 \cdot \left(P_1 + P_2\right)}/{p_3 \cdot \left(P_1 + P_2\right)}$. In the c.m. frame this corresponds to the fraction of the heavy-quark energy carried by the meson, $z' = E_{\rm meson}/E_{Q}$. The problem of this definition is best visible when $Y=y=0$, i.e. $z' = M_{\rm T}/m_{\rm T}$. The fragmentation functions are zero for $z' \geq 1$, which means that the partonic $p_{\rm T}$ has a lower limit $p_{\rm T}^2 \geq {P^2_{\rm T}+M^2-m^2} \geq {M^2-m^2}$. In other words, heavy quarks at sufficiently low transverse momenta will not form heavy-flavoured mesons at all. The definition of Eq.~(\ref{eq:fragvar}) evades this problem but also other choices are possible \cite{Albino:2008fy,Kniehl:2015fla}. An issue like this admittedly falls outside the predictive power of collinear factorization and can be categorized as modeling the higher-twist effects associated with the hadronization. We have checked that for the results presented in the present paper, the differences between the two above versions of the fragmentation variable, $z$ and $z'$, remains at most $\sim 10\%$ at small values of $p_{\mathrm{T}}$ -- well below the uncertainties originating e.g. from the scale choices -- and vanish completely at larger values of $p_{\mathrm{T}}$.

The partonic cross sections $d\hat\sigma^{ij\rightarrow k+X}$ in GM-VFNS are subject to a scheme dependence \cite{Thorne:2008xf} to accomplish a description valid at any $p_{\mathrm{T}}$. In the SACOT-$m_{\rm T}$ scheme \cite{Helenius:2018uul} the processes in which the heavy quarks are explicitly produced from massless flavours,
\begin{align}
& gg \rightarrow Q + X \,, \ \ qg \rightarrow Q + X \,, \ \ qq \rightarrow Q + X \,, \nonumber
\end{align}
are evaluated with partonic cross sections carrying the full heavy-quark mass dependence \cite{Nason:1989zy}, renormalized in the $\overline{\rm MS}$ scheme (see Sect.~3 of Ref.~\cite{Cacciari:1998it}). We will refer to these channels as being the ``direct'' ones. These fixed-order NLO cross sections contain logarithmic terms $\log \rho$ which originate from (i) collinear radiation of gluons off a final-state heavy quark, (ii) collinear splitting of final-state gluons into a heavy quark-antiquark pair, and (iii) collinear splitting of initial-state gluons into a pair of heavy quark and antiquark. These logarithms can be resummed into the scale dependence of the heavy-quark PDFs $f_Q(x,\mu^2_{\rm fact})$ and parton-to-meson FFs, $D_{k \rightarrow h_3}(z,\mu^2_{\rm frag})$. The resummation then gives rise to the contributions (i) with heavy quarks in the initial state and (ii) in which the fragmentation is initiated by a light parton. We will refer to these channels as being the ``non-direct'' ones. In our scheme, these processes are evaluated with the zero-mass (ZM) $\overline{\rm MS}$ expressions for the partonic cross sections $d\hat\sigma^{ij\rightarrow k+X}(\tau_1^0, \tau_2^0)|_{\rm ZM}$ \cite{Aversa:1988vb}, where
\begin{align}
\tau_1^0  = \frac{p_{\rm T}e^{-y}}{x_2 \sqrt{s}} 
\,, \ \ \ \
\tau_2^0  = \frac{p_{\rm T}e^{y}}{x_1 \sqrt{s}} 
\,, \ \ \ \
\end{align}
but replacing the massless variables $\tau_1^0$ and $\tau_2^0$ by the massive invariants $\tau_1$ and $\tau_2$ defined in Eq. (\ref{eq:tau12}). In summary,
\begin{align}
& \hspace{-4.2cm} {\rm \bf Direct \hspace{-0.1cm}:} \nonumber \\
\frac{\mathrm{d}^3\hat{\sigma}^{ij\rightarrow k+X}}{\mathrm{d}^2p_{\rm T} \mathrm{d}y}\bigg|_{{\rm SACOT-}m_{\rm T}}
& \hspace{-0.5cm} \equiv
\frac{\mathrm{d}^3\hat{\sigma}^{ij\rightarrow k+X}(\tau_1, \tau_2, \rho)}{\mathrm{d}^2p_{\rm T} \mathrm{d}y} \,, \ \ \ \ \ \ \,
ij\rightarrow k + X \in \left\{
\begin{array}{c}
gg \rightarrow Q + X \\
qg \rightarrow Q + X \\
qq \rightarrow Q + X 
\end{array} \right.  \\
\nonumber \\
& \hspace{-4.2cm} {\rm \bf Non{\text -}direct \hspace{-0.1cm}:} \nonumber \\
\frac{\mathrm{d}^3\hat{\sigma}^{ij\rightarrow k+X}}{\mathrm{d}^2p_{\rm T} \mathrm{d}y}\bigg|_{{\rm SACOT-}m_{\rm T}}
& \hspace{-0.5cm} \equiv
\frac{\mathrm{d}^3\hat{\sigma}^{ij\rightarrow k+X}(\tau_1, \tau_2)}{\mathrm{d}^2p_{\rm T} \mathrm{d}y} \bigg|_{\rm ZM} \,, \ \ \ \ \ \
ij\rightarrow k + X \notin \left\{
\begin{array}{c}
gg \rightarrow Q + X \\
qg \rightarrow Q + X \\
qq \rightarrow Q + X 
\end{array} \right. \label{eq:defsacotmt2}
\end{align}
To motivate the latter choice we note that to (i) retain the Lorentz invariance, and (ii) recover the zero-mass $\overline{\rm MS}$ result in the $p_{\rm T} \rightarrow \infty$ limit, Eq.~(\ref{eq:defsacotmt2}) is a rather natural choice. It implicitly accounts for the fact that even in an apparently massless production channels like $gg \rightarrow g(\rightarrow Q\overline{Q}) + X$, the final-state parton will eventually split into a heavy quark-antiquark pair such that the relevant variables to describe the underlying process are the massive invariants $\tau_{1,2}$, not the massless ones $\tau^0_{1,2}$ to account for finite virtualities of the intermediate partons. This choice also ensures that the cross sections remain finite in the $p_{\rm T} \rightarrow 0$ limit. 

The subtractions in Eq.~(\ref{eq:master}) associated with the initial-state radiation are obtained by replacing the heavy-quark PDFs $f_{Q}(x,\mu^2_{\rm fact})$ by
\begin{align}
f_{Q}(x,\mu^2_{\rm fact}) & \longrightarrow  \bigg(\frac{\alpha_s}{2\pi} \bigg) \log\left( \frac{\mu^2_{\rm fact}}{m^2}\right) \int_x^1 \frac{dz}{z} P_{qg}\left(\frac{x}{z}\right) f_g(z,\mu^2_{\rm fact}) \\ 
P_{qg}(z) & = \frac{1}{2} \Big[z^2 + (1-z)^2 \Big]
\end{align}
in the 
\begin{align}
& Qg \rightarrow Q + X \,, \ \ Qq \rightarrow Q + X \,, \nonumber
\end{align}
channels, and keeping terms up to $\alpha_s^3$. Similarly, the subtractions associated with the final-state radiation are obtained by replacing the FFs by
\begin{align}
D_{Q \rightarrow h_3}(x,\mu^2_{\rm frag}) & \longrightarrow \bigg(\frac{\alpha_s}{2\pi} \bigg)  \int_x^1 \frac{dz}{z}
d_{QQ}\left(\frac{x}{z} \right) D_{Q \rightarrow h_3}(z,\mu^2_{\rm frag}) \,,
 \\
D_{g \rightarrow h_3}(x,\mu^2_{\rm frag}) & \longrightarrow \bigg(\frac{\alpha_s}{2\pi} \bigg)  \log\left( \frac{\mu^2_{\rm frag}}{m^2}\right) \int_x^1 \frac{dz}{z} 
P_{qg}\left(\frac{x}{z}\right) D_{Q \rightarrow h_3}(z,\mu^2_{\rm frag}) \,,  \\
d_{QQ}(z) & = C_f \left\{\frac{1+z^2}{1-z}\left[\log\left( \frac{\mu^2_{\rm frag}}{m^2}\right) -2\log(1-z) - 1 \right] \right\}_+ \,,
\end{align}
in the 
\begin{align}
& gg \rightarrow Q + X \,, \ \ qg \rightarrow Q + X \,, \ \ qq \rightarrow Q + X \,, \nonumber \\
& gg \rightarrow g + X \,, \ \ \ qg \rightarrow g + X \,, \ \ \ qq \rightarrow g + X \,, \nonumber 
\end{align}
channels, and keeping terms up to $\alpha_s^3$. The non-logarithmic terms in $d_{\rm QQ}$ are associated with the definition of the $\overline{\rm MS}$ FFs in the presence of a finite quark mass \cite{Mele:1990cw,Melnikov:2004bm}. In addition, the fully massive calculation used to evaluate the direct contributions \cite{Nason:1989zy} are renormalized in the so-called decoupling scheme \cite{Collins:1978wz} in which the scale dependence of $\alpha_s$ excludes the contributions from heavy-quark loops. To translate the results in the decoupling scheme to the usual $\overline{\rm MS}$ scheme, additional terms are supplied, see Sect.~3 of Ref.~\cite{Cacciari:1998it}.

There are three independent scales involved in our calculation -- the renormalization, factorization and fragmentation scales. These are taken to be
\begin{align}
\mu_i =  c_i \sqrt{P_{\rm T}^2 + m^{2}} \,,
\end{align}
where $m$ is the heavy-quark mass and our default choice is $c_i=1$, as in Ref.~\cite{Helenius:2018uul}. To chart the dependence of our results on this choice we repeat the calculations by taking $c_i = 0.5, 1, 2$, with a restriction,
\begin{align}
\frac{1}{2} \leq \frac{\mu_{\rm ren}}{\mu_{\rm fact}} \leq 2 \,, \ \ \ \frac{1}{2} \leq \frac{\mu_{\rm ren}}{\mu_{\rm frag}} \leq 2 \,. 
\end{align}
In total there are then 17 different scale combinations whose envelope we take as the scale uncertainty. We note that the FFs become scale independent for $\mu_{\rm frag} \leq m$ and only $D_{Q \rightarrow h_3}$ is non-zero in this regime. The heavy-quark PDFs are zero for $\mu_{\rm fact} \leq m$. Consistently, no initial-state subtraction terms are included when $\mu_{\rm fact} \leq m$, and no final-state subtraction terms are included when $\mu_{\rm frag} \leq m$.

We take the B-meson FFs from the Kniehl-Kramer-Schienbein-Spiesberger analysis (KKSS08) \cite{Kniehl:2007erq} which fits the SLD \cite{SLD:2002poq}, OPAL \cite{OPAL:2002plk}, and ALEPH \cite{ALEPH:2001pfo} data on B-meson production in $e^+e^-$ annihilation near the Z-boson pole, $\sqrt{s} = M_{\rm Z}$. Recently, also FFs at NNLO accuracy have become available \cite{Czakon:2022pyz}. It should be noted that in the KKSS08 analysis the bottom mass was taken to be $m_{\rm b, FF}=4.5 \, {\rm GeV}$, which differs from the values employed in the PDFs we will use in our calculations, $m_{\rm b}=4.92 \, {\rm GeV}$ for NNPDF4 \cite{NNPDF:2021njg}, and $m_{\rm b}=4.75 \, {\rm GeV}$ for MSHT20 \cite{Bailey:2020ooq}. Since the data in KKSS08 analysis are taken at $\sqrt{s} = M_{\rm Z}$, the exact value of the bottom-quark mass used there cannot be very critical. On the other hand, the PDF fits utilize much more data at lower interaction scales and are thus arguably much more sensitive to the quark masses (i.e. changing the input masses changes the PDFs). We thus find it better justified to use the mass values from PDFs in our calculations. To ensure the correct behaviour at the threshold, the final-state subtraction terms as well as e.g. the gluon FFs should vanish at the scale $\mu_{\rm frag} = m_{\rm b}$. To enforce this, we always use the scale  $\mu^2 = c_i \sqrt{P_{\rm T}^2 + m^{2}_{\rm b, FF}}$ when calling the FFs. In the future, to avoid making such compromises, it would be useful to have the B-meson FFs available with the exact mass values utilized in the global PDF fits. In most of our calculations we will use the NNPDF4 partons \cite{NNPDF:2021njg} which constitute the most recent set. To investigate the mass dependence, we will use a special version of MSHT20,  \textsc{MSHT20nlo\_mbrange\_nf5} \cite{Cridge:2021qfd} which provides PDF fits with different bottom-quark masses including also the one which matches with $m_{\rm b, FF}$. In the case of p-Pb collisions, we will use the EPPS21 nuclear PDFs \cite{Eskola:2021nhw} (with CT18A baseline proton PDFs  \cite{Hou:2019efy}) for which $m_{\rm b}=4.75 \, {\rm GeV}$, and the nNNPDF3.0 nuclear PDFs \cite{Khalek:2022zqe} for which $m_{\rm b}=4.92 \, {\rm GeV}$.

\begin{figure}[htb!]
\centering
\includegraphics[width=0.75\linewidth]{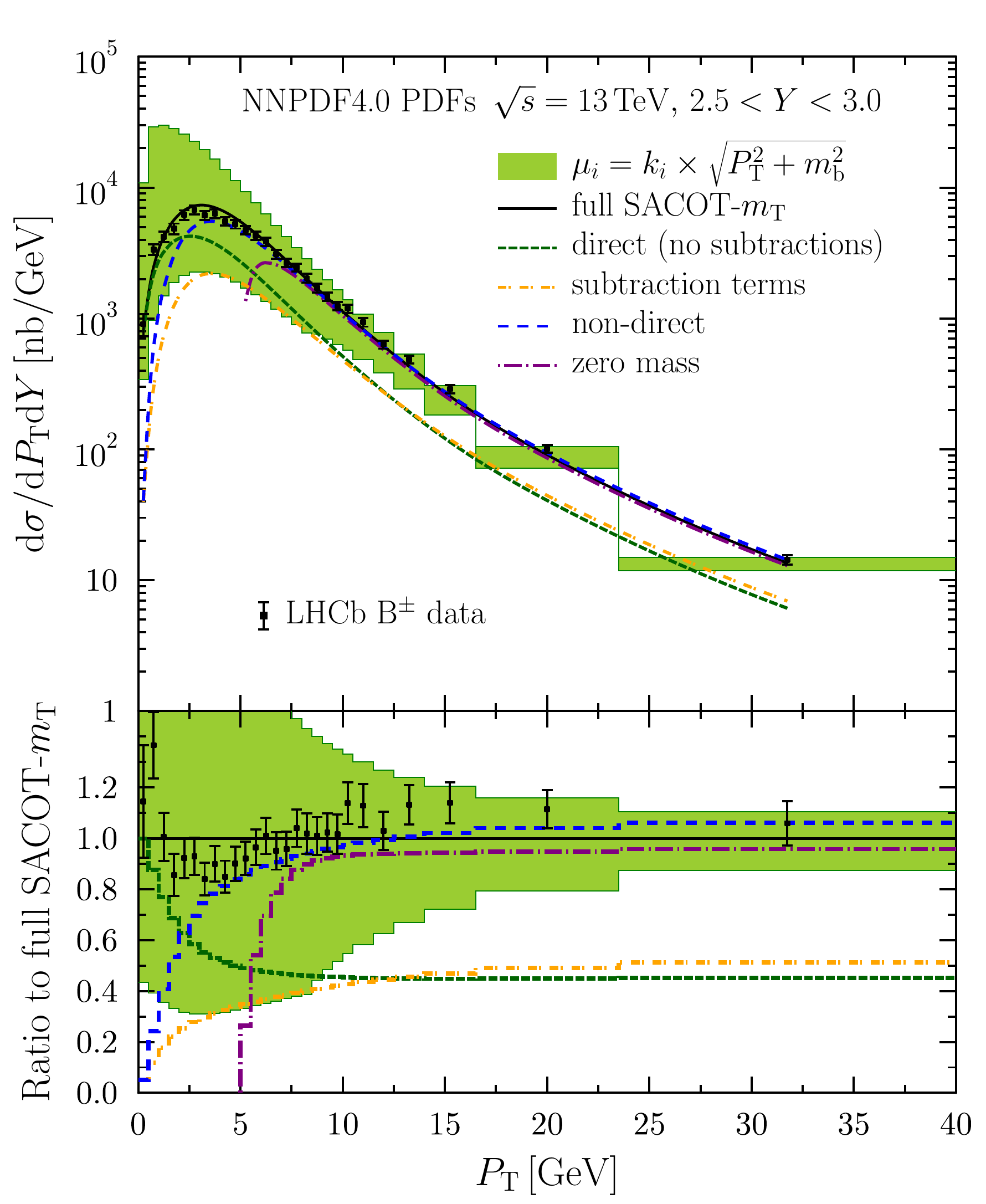}
\caption{
The $13\,{\rm TeV}$ B$^\pm$-meson data of the LHCb collaboration \cite{LHCb:2017vec} in the rapidity window $2.5 < Y < 3.0$ compared with the SACOT-$m_{\rm T}$ calculation. The plot shows separately the full calculation (black solid), the direct i.e. light-parton to heavy quark production channels (green dashed), subtraction terms (yellow dotted-dashed), non-direct production channels (blue dashed), and the zero-mass calculation (purple dotted-dashed). The filled bands correspond to the uncertainty from the scale variation.
}
\label{fig:diag1}
\end{figure} 

\section{Results for proton-proton collisions}
\label{sec:results-pp}

To highlight the key features of the SACOT-$m_{\mathrm{T}}$ setup we present, in Figure~\ref{fig:diag1}, the B$^\pm$ cross sections at $\sqrt{s} = 13 \, {\rm TeV}$ in the rapidity window $2.5 < Y < 3.0$ together with the experimental data from the LHCb collaboration \cite{LHCb:2017vec}. The full SACOT-$m_{\rm T}$ calculation follows the data very well and, in particular, reproduces the turnover at $P_{\rm T} \approx 3 \, {\rm GeV}$. The scale uncertainty is shown as the green band which is large at small values of $P_{\mathrm{T}}$ but reduces to 10\% at highest considered values of $P_{\mathrm{T}}$. To illustrate how the B-meson cross section in our scheme builds from various components, Figure~\ref{fig:diag1} also shows separately the contributions of direct terms, subtraction terms, and the non-direct parts in which there are either bottom quark(s) in the initial state or in which the fragmentation proceeds from a light parton. At low $P_{\rm T}$ the direct part clearly dominates and, by construction, is the only contribution at $P_{\rm T} = 0~\text{GeV}$. As $P_{\rm T}$ increases the subtraction terms approximate rather well the full direct contribution and eventually the net contribution of these two becomes rather small. When this happens, the contributions from initial-state heavy quarks and light-parton fragmentation are the dominant ones. With our default choice of scales, this begins to happen already around $P_{\rm T} \approx m_{\mathrm{b}}$. Arguably, the collinear logarithms $\sim \log\big(p_{\rm T}^2/m^2 \big)$ are not yet particularly large at such values of $P_{\rm T}$ so their resummation should not be a too big of an effect either. However, even if the resummation would not yet be a large effect, the non-direct contributions can be significant as e.g. the $gg \rightarrow gg$ matrix element that enters the contribution from gluon fragmentation carries a large colour factor which increases its importance even if the associated logarithm would not yet be particularly large. For $P_{\rm T} \approx m_{\mathrm{b}}$ and higher, the full calculation is significantly above the direct part. On one hand this is due to the $\alpha_s^3$ terms in the contributions with initial-state heavy quarks or light-parton fragmentation, which also partly catch the NNLO contributions to the fixed-order calculation which are now known to be important \cite{Catani:2020kkl}. On the other hand, towards higher values of $P_{\rm T}$ the resummation of the collinear logarithms becomes also gradually more important of an effect. The scale variations result in a significant uncertainty band. Part of this largeness is related to the fact that the scale choice also controls the relative importance of the direct, subtraction, and non-direct contributions. For example, with $k_i=1/2$ only the direct part contributes up to $P_{\rm T} = \sqrt{3} m_{\mathrm{b}} \approx 8.5 \, {\rm GeV}$, whereas with our default choice of scales it is the non-direct part that clearly dominates at $P_{\rm T} = 8.5 \, {\rm GeV}$. The result of a fully zero-mass calculation, but still adopting our default choice of scales, is shown in Figure~\ref{fig:diag1} as well. We see that the zero-mass calculation agrees rather well with the full SACOT-$m_{\rm T}$ result already at $P_{\rm T} \gtrsim 2m_{\mathrm{b}}$ though the residual mass effects die out rather slowly in $P_{\rm T}$. Towards lower values of $P_{\rm T}$ the NLO zero-mass cross section not only diverges but goes also negative due the spurious behaviour of the zero-mass coefficient functions. 

\begin{figure}[htb!]
\centering
\includegraphics[width=0.75\linewidth]{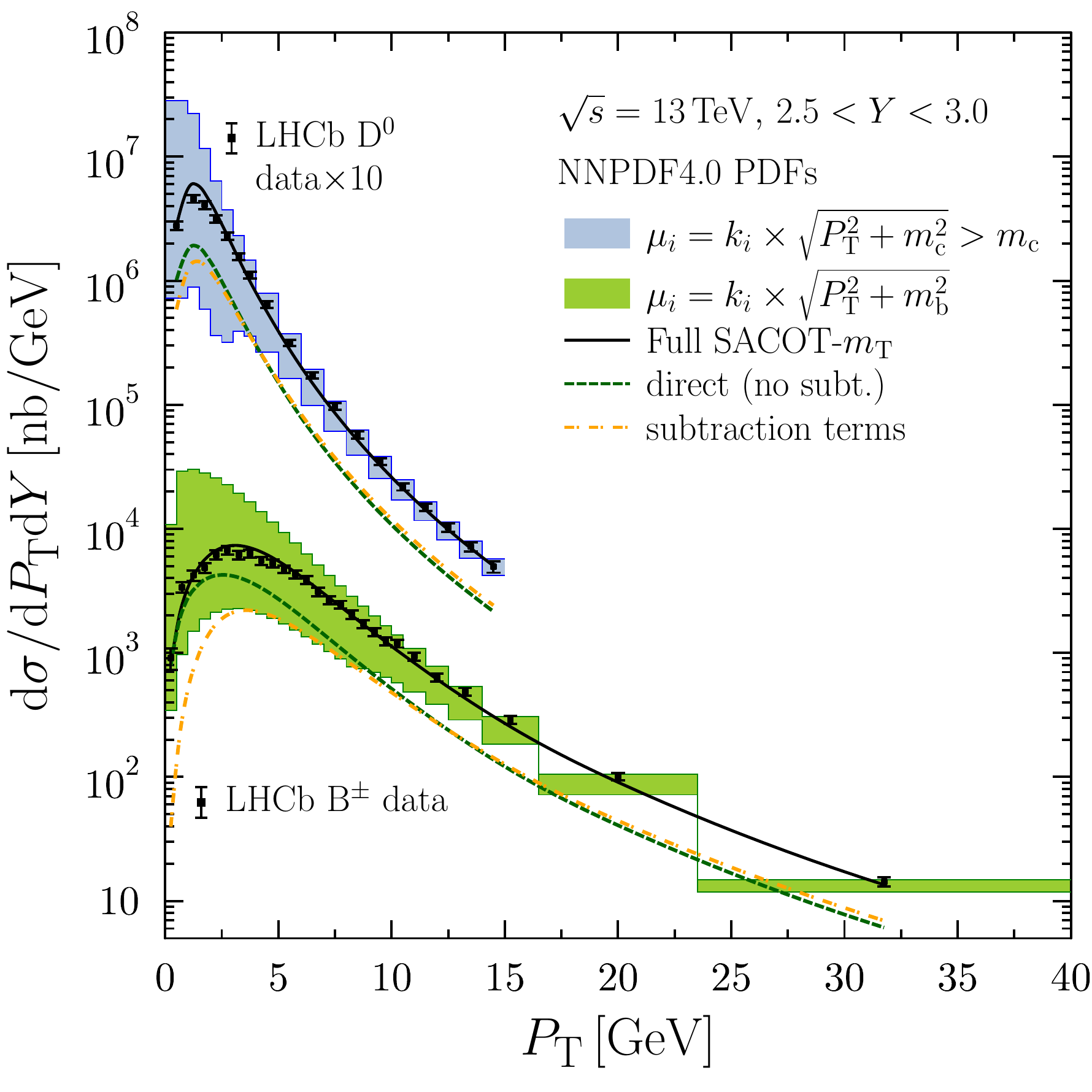}
\caption{Comparison of D$^0$- (upper curves) and B$^\pm$-meson (lower curves) production in the SACOT-$m_{\rm T}$ scheme at $\sqrt{s} = 13\,{\rm TeV}$, $2.5 < Y < 3.0$. Black solid curves correspond to the full calculation and the green dashed curves to the contributions from the direct (i.e. light parton to heavy quark) production channels. The orange dashed-dotted curves are the subtraction terms and the filled bands show the uncertainty from the scale variation. The data are from the LHCb collaboration \cite{LHCb:2015swx,LHCb:2017vec}. For clarity, the data and curves corresponding to the D$^0$ mesons have been multiplied by a factor of 10.}
\label{fig:BvsD}
\end{figure} 

\begin{figure}[htb!]
\centering
\includegraphics[width=0.73\linewidth]{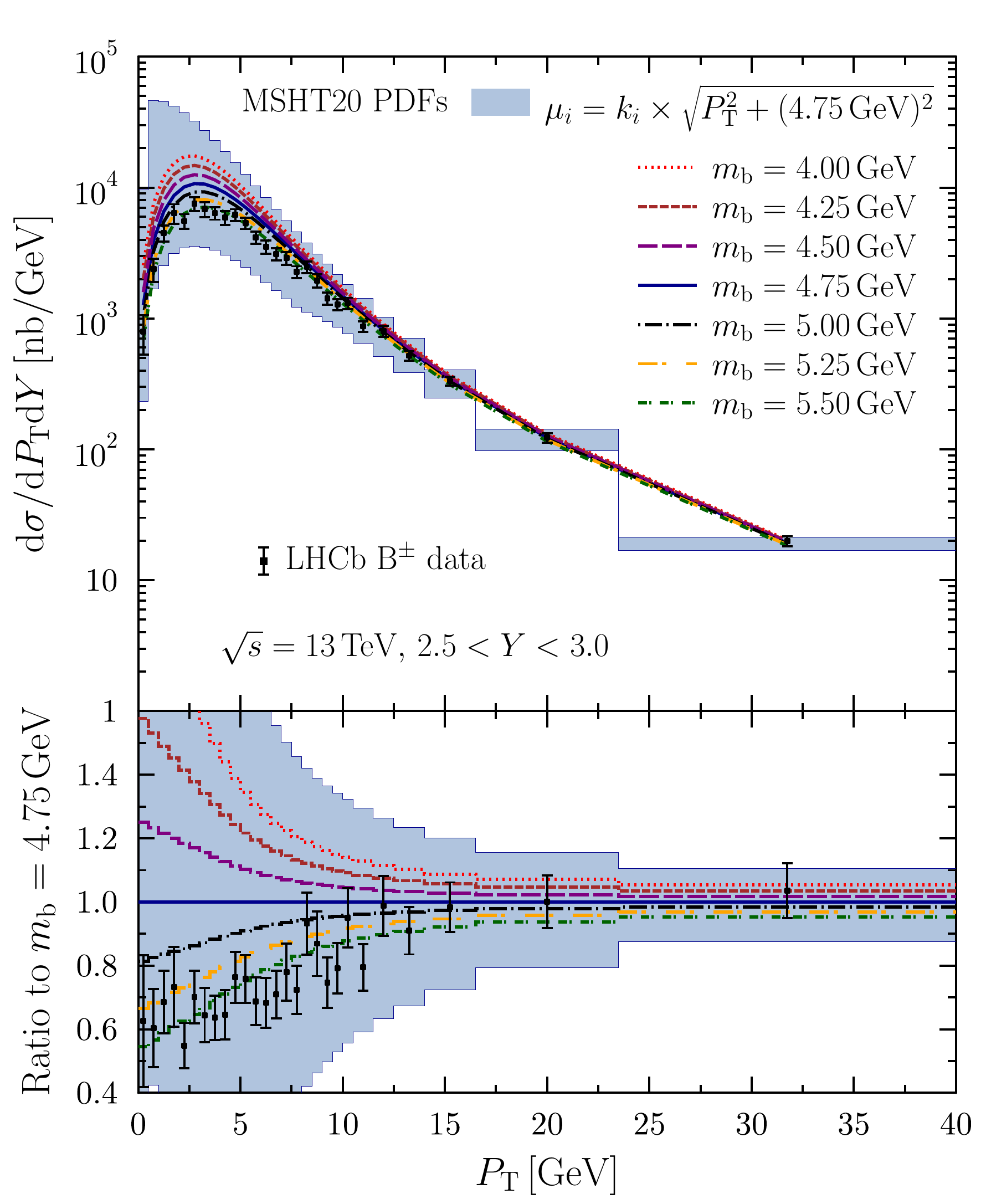}
\caption{
The $13\,{\rm TeV}$ B$^\pm$-meson data of the LHCb collaboration \cite{LHCb:2017vec} in the rapidity window $2.5 < Y < 3.0$ compared with the SACOT-$m_{\rm T}$ calculation with different bottom-quark masses $m_{\mathrm{b}}$. The calculation uses the \textsc{MSHT20nlo\_mbrange\_nf5} partons \cite{Cridge:2021qfd} which are available for $m_{\mathrm{b}} = 4.0\ldots5.50 \,{\rm GeV}$. The scale uncertainty band was evaluated with $m_{\mathrm{b}} = 4.75 \,{\rm GeV}$. 
}   
\label{fig:differentPDFs}
\end{figure} 

The observations made above are reminiscent of those we found earlier for D mesons \cite{Helenius:2018uul} but the effects of heavy-quark mass simply persist up to higher values of $P_{\rm T}$. This is illustrated in Figure~\ref{fig:BvsD} where we plot the D$^0$ results in the same figure. For the D-meson data, the turnover happens at lower $P_{\rm T}$ in comparison to the B-meson case. This behaviour is also well reproduced by our default scale choice -- a larger quark mass more strongly ``screens'' the partonic propagators due to larger virtuality and shifts the turnover to larger $P_{\rm T}$. One can also clearly see that -- in our scheme and the default choice of scales -- the subtraction terms approximate well the contributions from the direct production channels for D mesons immediately above zero $P_{\rm T}$, whereas for B mesons the cancellation between the subtraction terms and the direct production channels is shifted to larger $P_{\rm T}$. In comparison to the D-meson results, and perhaps a little surprisingly, the scale uncertainty remains larger for B mesons up to higher $P_{\rm T}$ although the associated QCD scales are larger. The reason is that for the B-meson production the interplay between various components (direct, non-direct, subtraction) remains non-trivial up to higher values of $P_{\rm T}$ and the dependence of this interplay on the scale choices results in a larger scale uncertainty. In the case of D mesons, the non-direct components quickly dominate the cross section with all considered scale choices. Notice that here we have limited the scales from below by the charm mass to make sure that they stay above the initial scales of the PDF analyses.

\begin{figure}[htb!]
\vspace{0.0cm}
\centering
\includegraphics[width=0.495\linewidth]{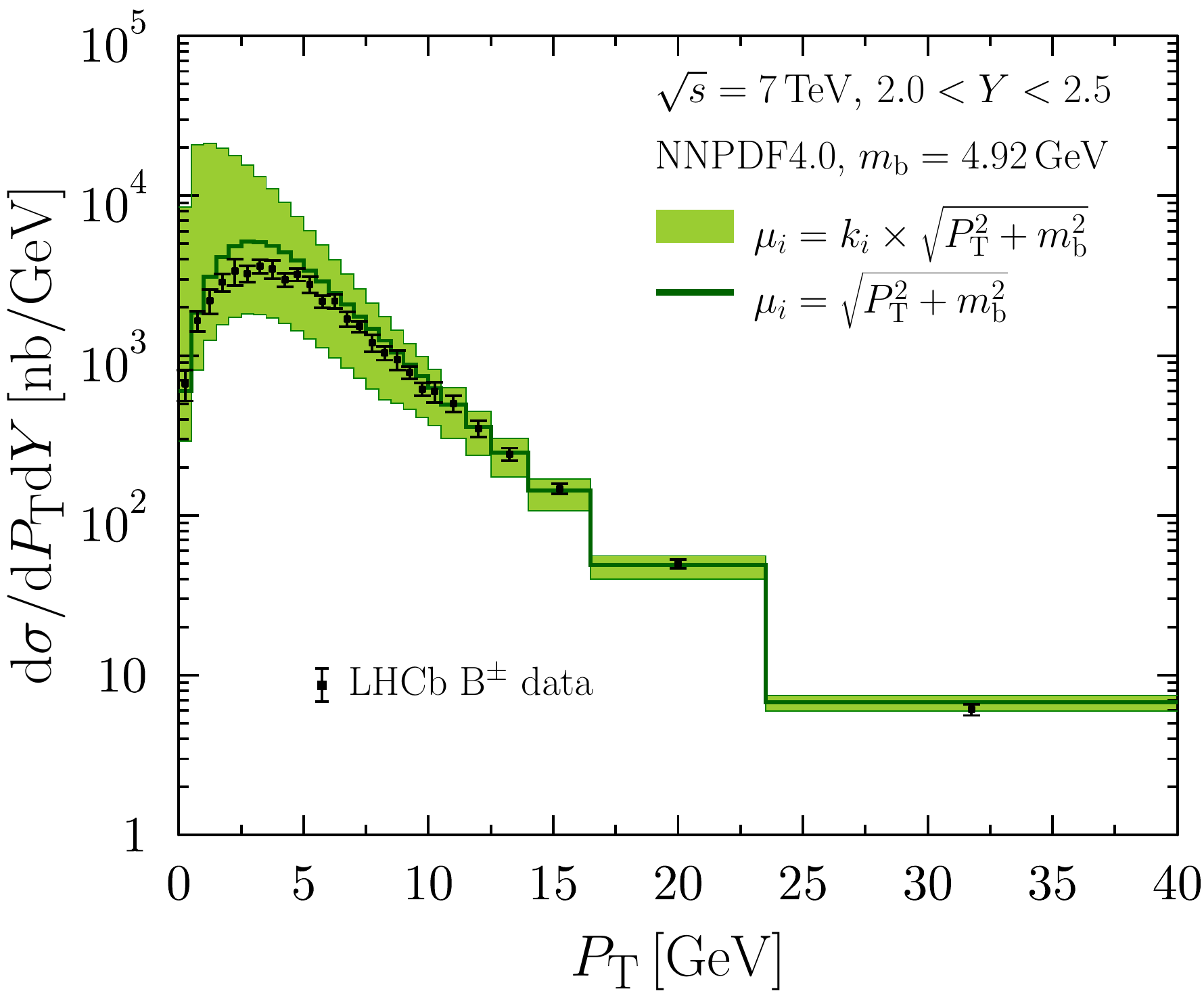}
\includegraphics[width=0.495\linewidth]{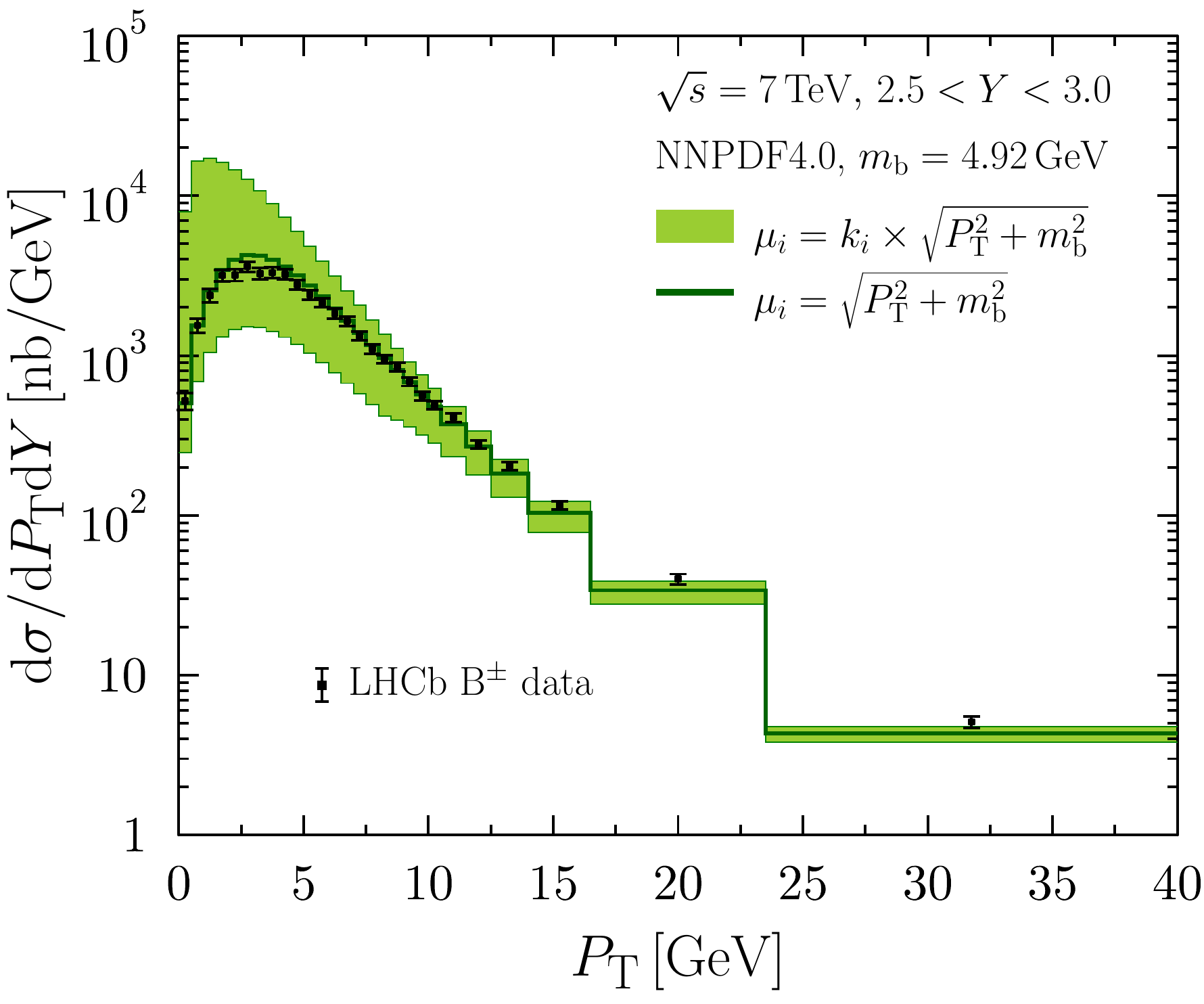} 
\includegraphics[width=0.495\linewidth]{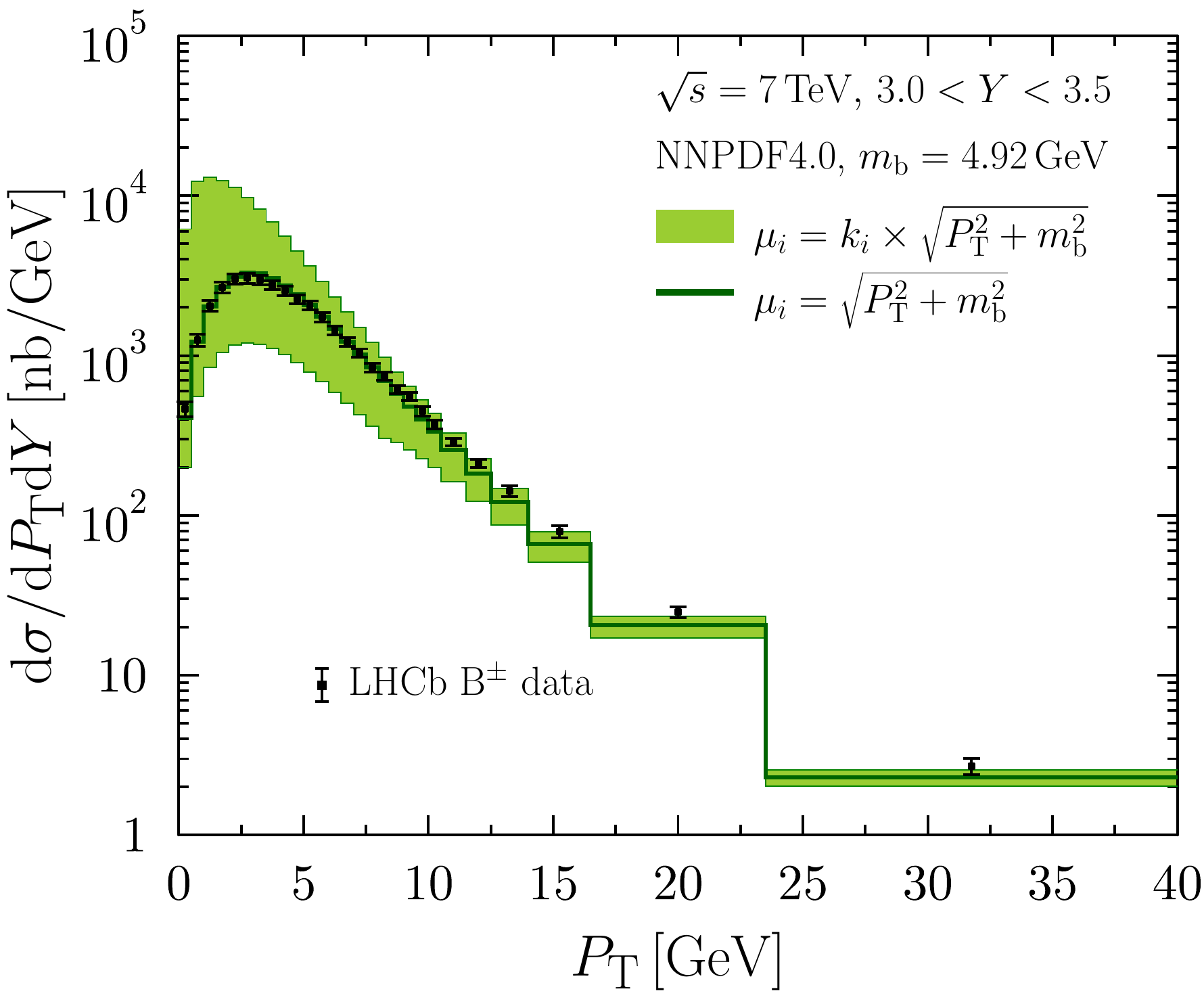}
\includegraphics[width=0.495\linewidth]{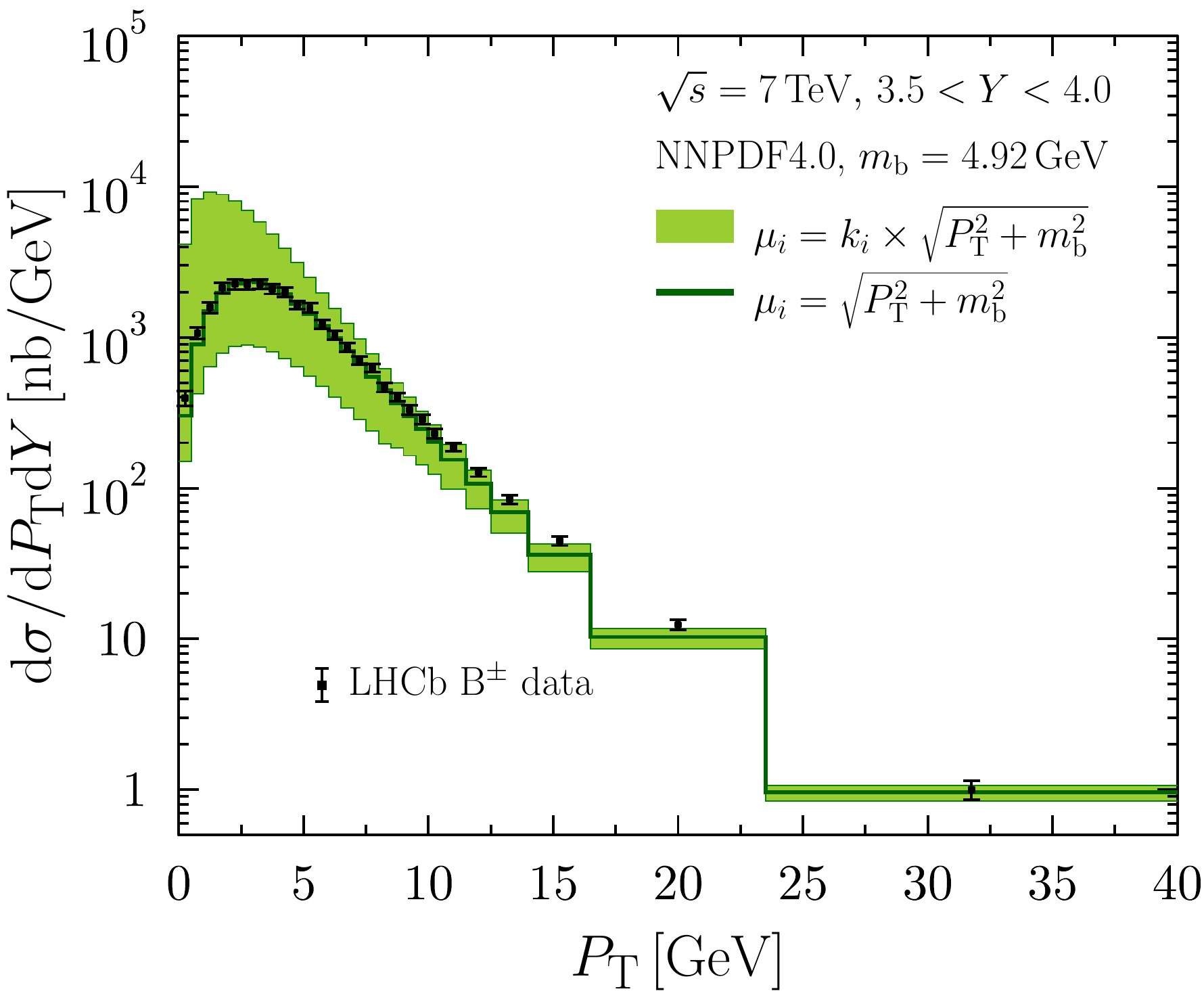}
\includegraphics[width=0.495\linewidth]{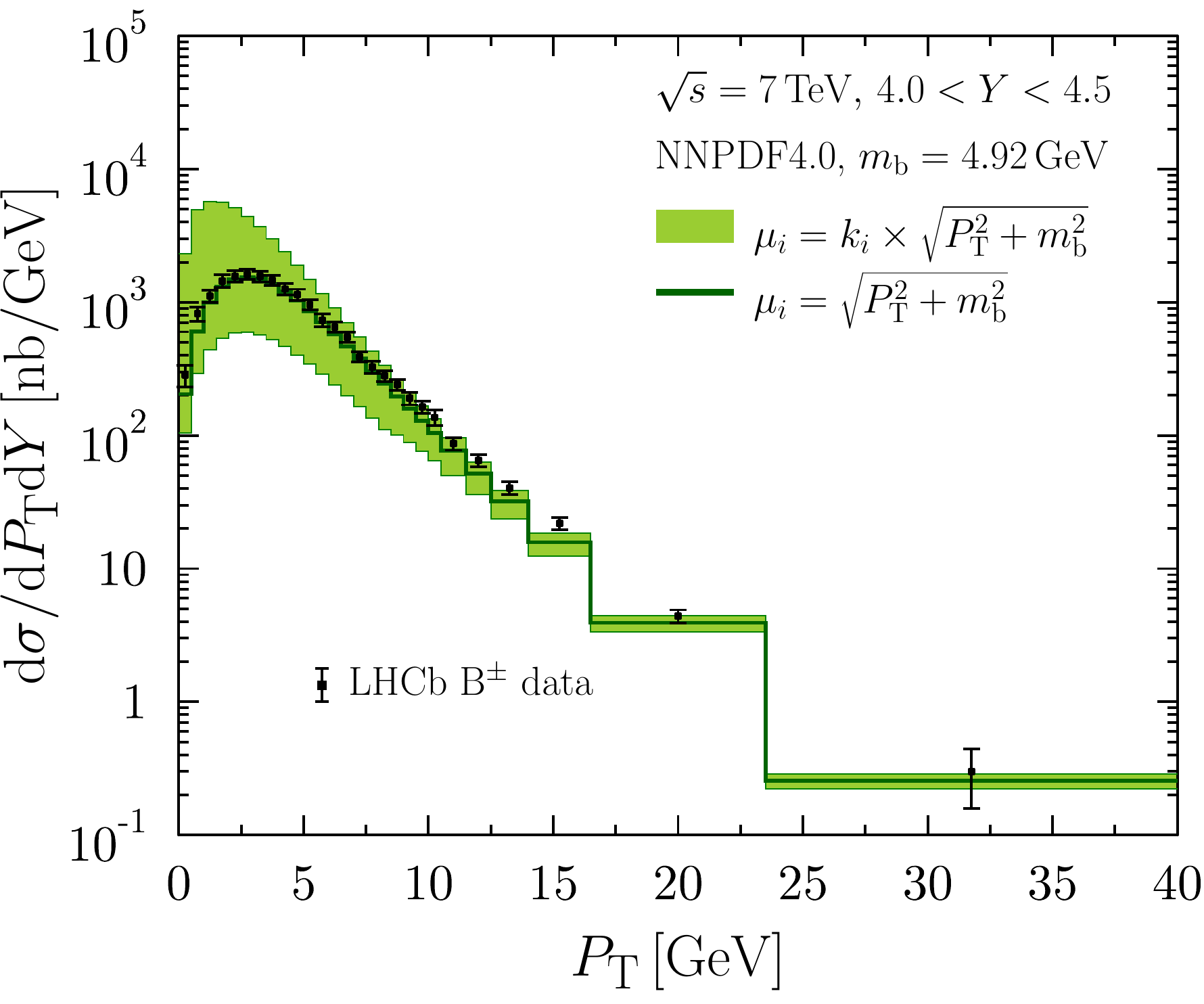}
\vspace{0.0cm} \caption{
The $7\,{\rm TeV}$ B$^\pm$-meson data of the LHCb collaboration \cite{LHCb:2017vec} compared with the SACOT-$m_{\rm T}$ calculation. Each panel correspond to a different rapidity window. The green solid curves show the results of our central scale choice $\mu_i = \sqrt{P_{\rm T}^2 + m_{\mathrm{b}}^2}$, and the light-green filled bands correspond to the uncertainty due to the scale variations.
}
\label{fig:LhCb7TeV}
\end{figure} 

\begin{figure}[htb!]
\centering
\includegraphics[width=0.495\linewidth]{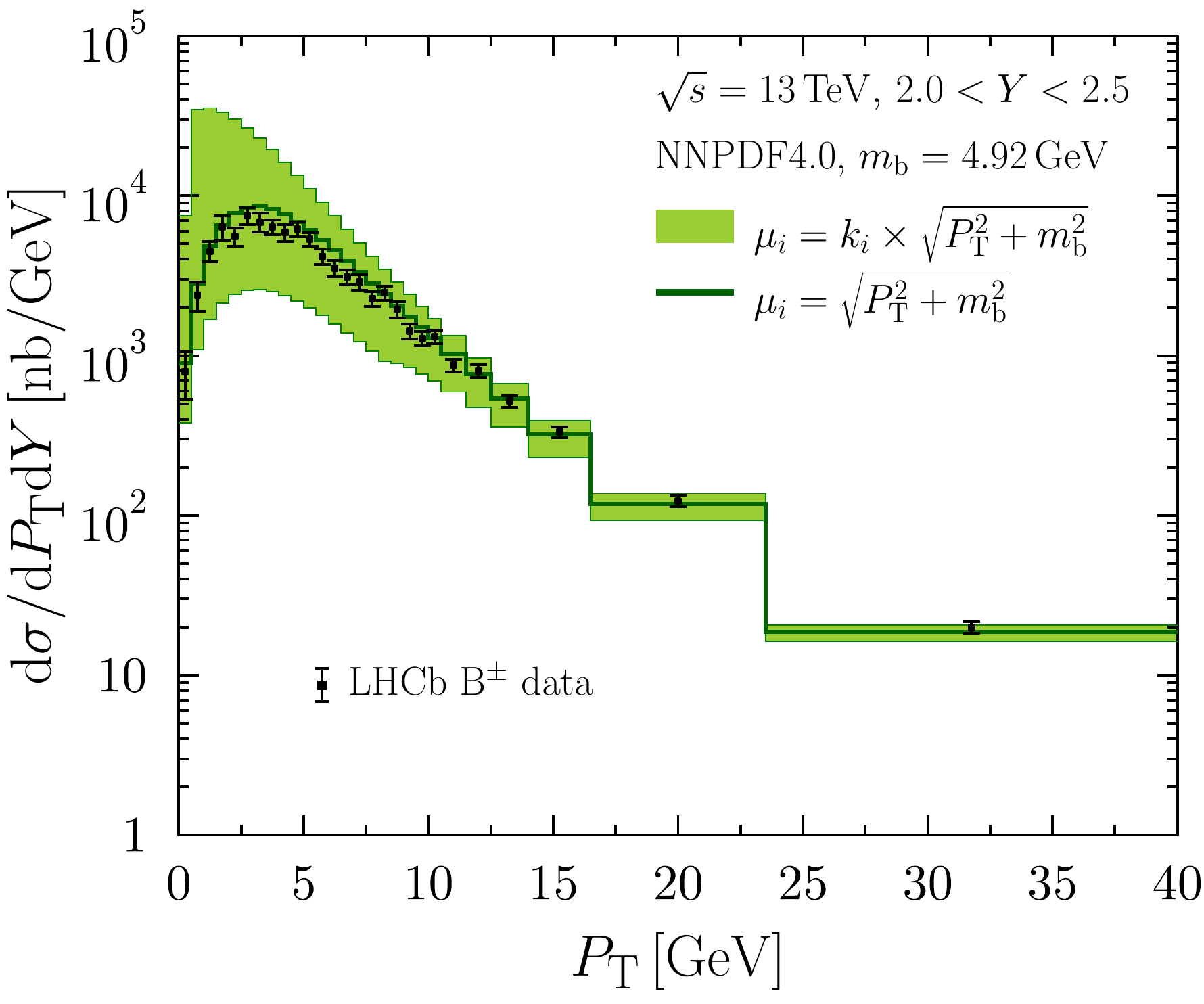}
\includegraphics[width=0.495\linewidth]{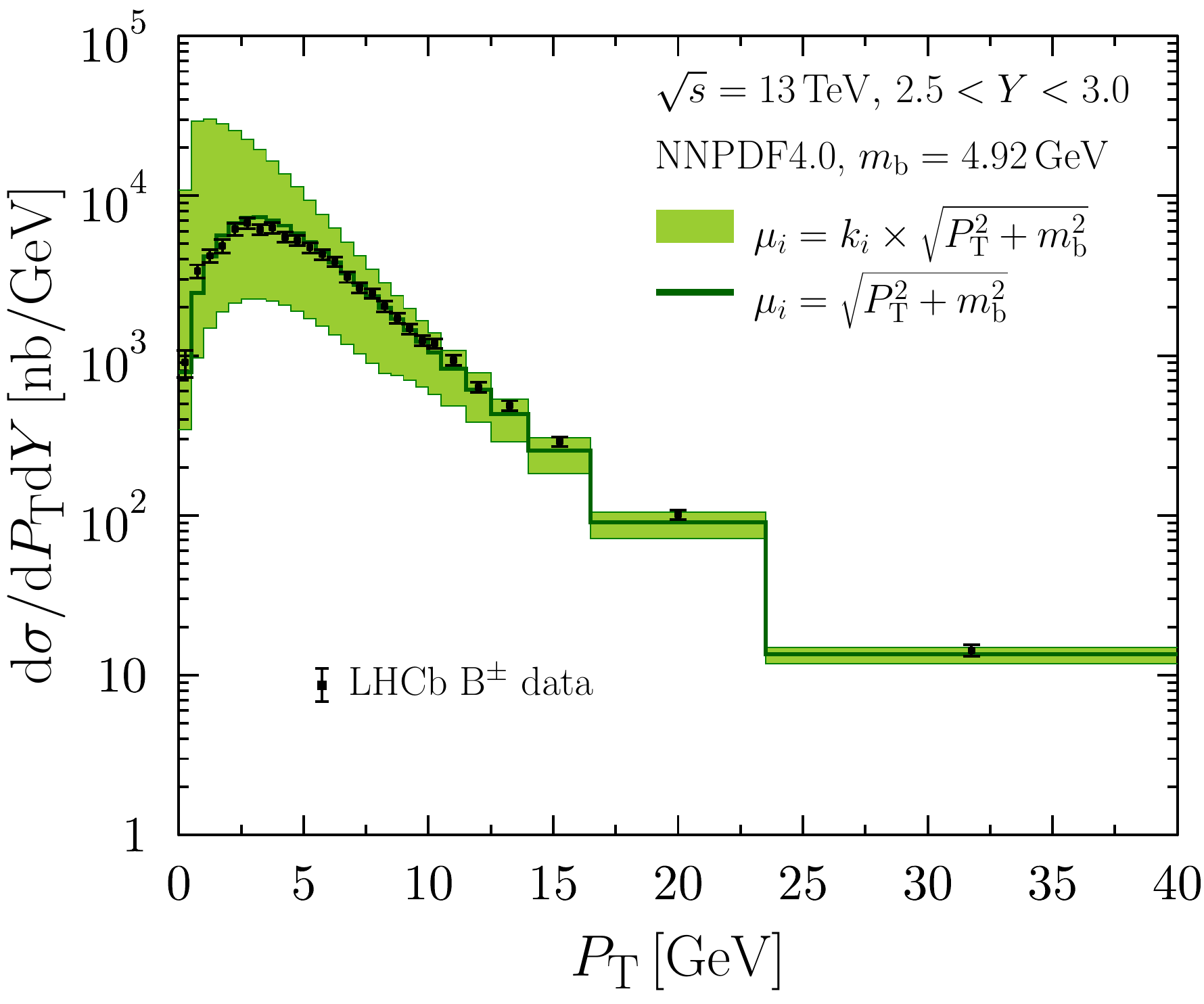}
\includegraphics[width=0.495\linewidth]{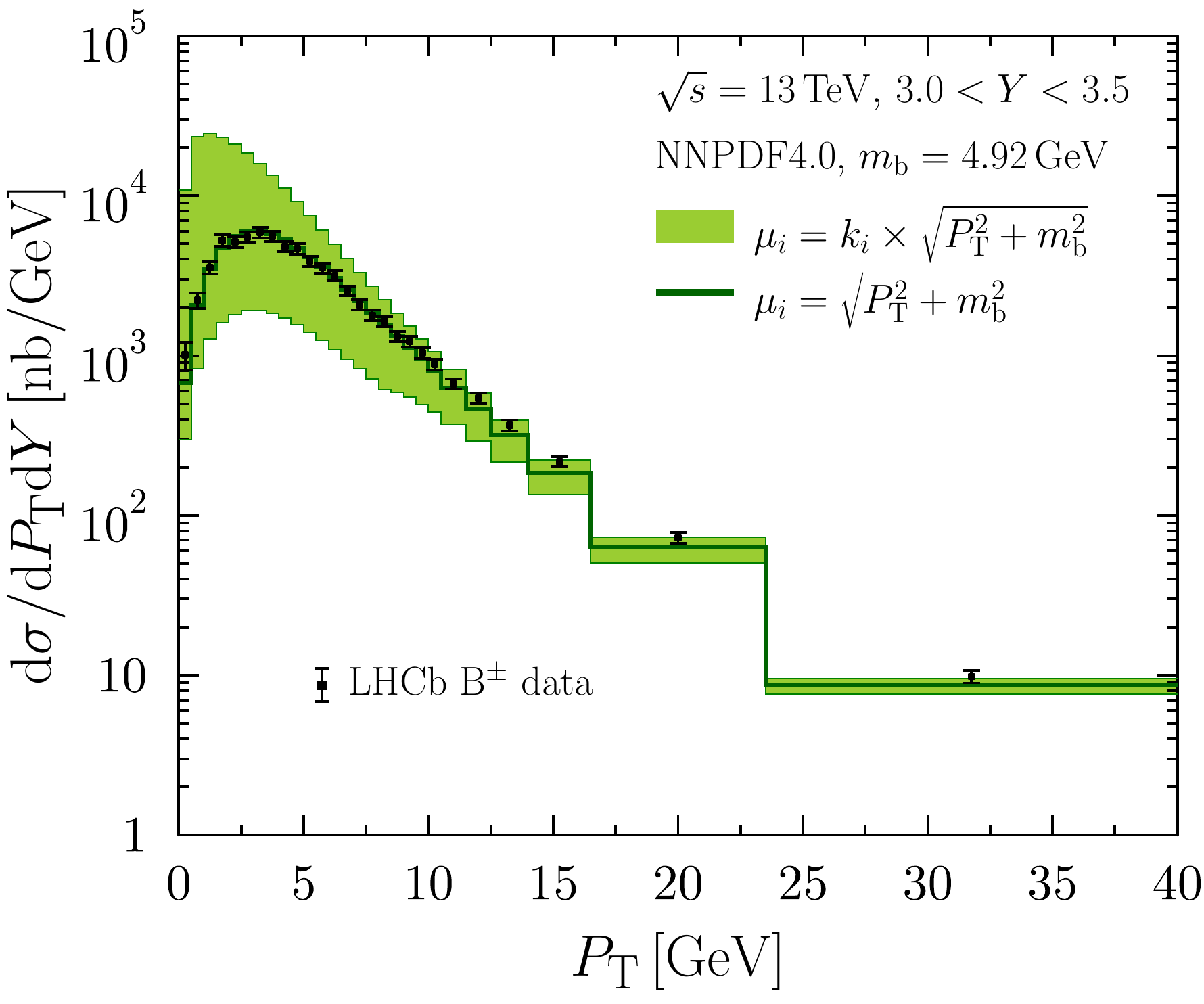}
\includegraphics[width=0.495\linewidth]{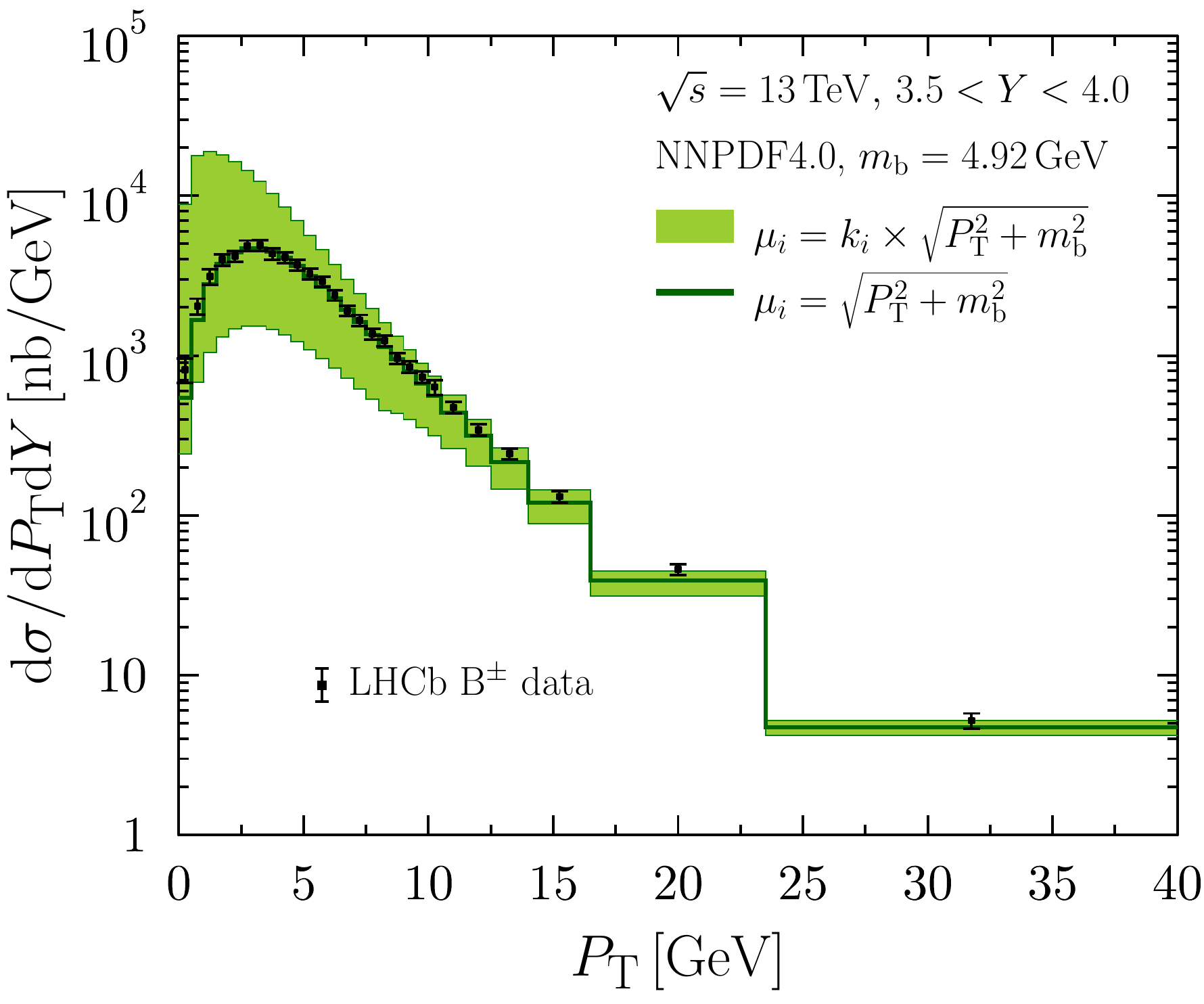}
\includegraphics[width=0.495\linewidth]{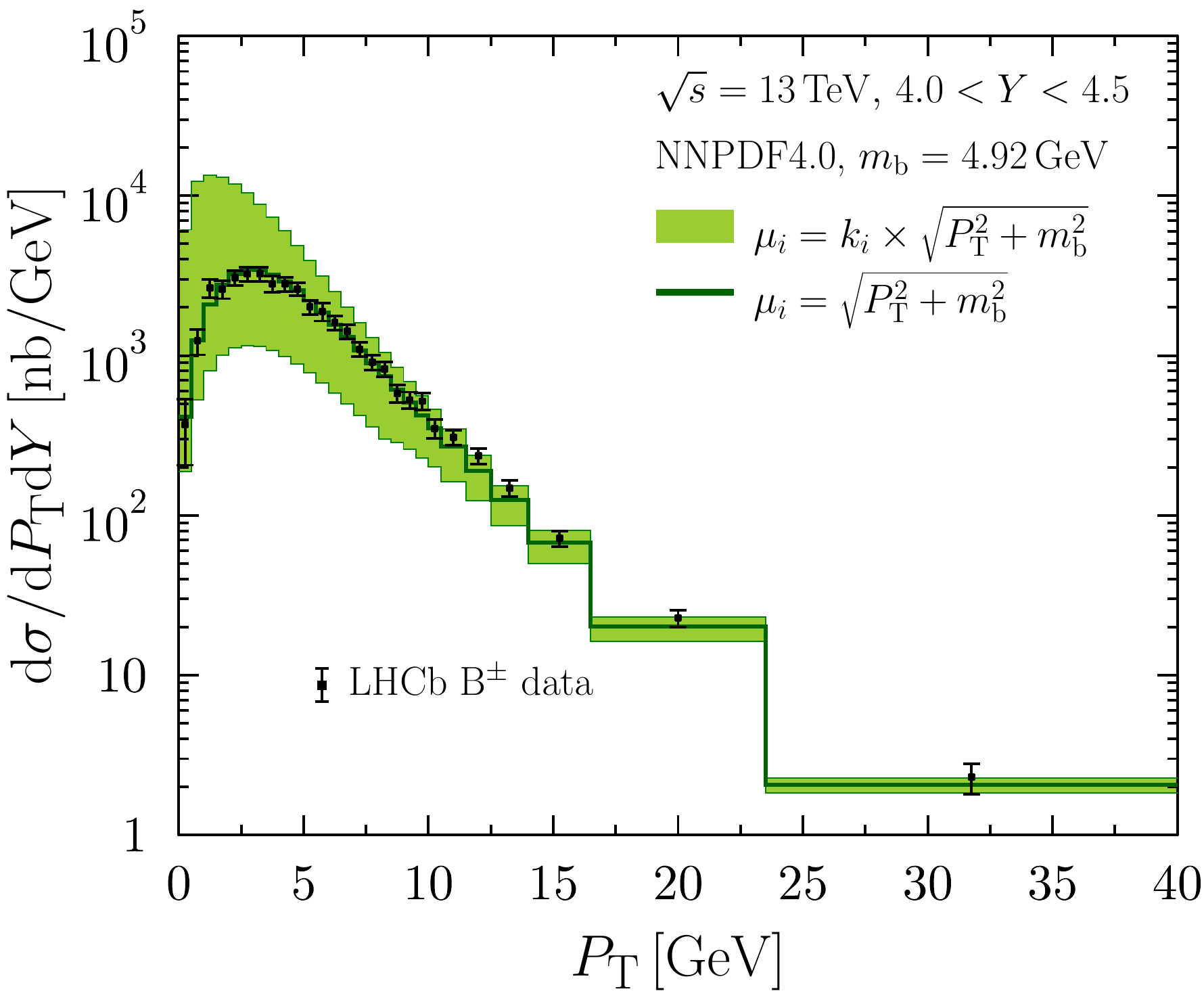}
\caption{
The $13\,{\rm TeV}$ B$^\pm$-meson data of the LHCb collaboration \cite{LHCb:2017vec} compared with the SACOT-$m_{\rm T}$ calculation. Each panel correspond to a different rapidity window. The green solid curves show the results of our central scale choice $\mu_i = \sqrt{P_{\rm T}^2 + m_{\mathrm{b}}^2}$, and the light-green filled bands correspond to the uncertainty due to the scale variations.
}
\label{fig:LhCb13TeV}
\end{figure} 

\begin{figure}[htb!]
\centering
\includegraphics[width=0.495\linewidth]{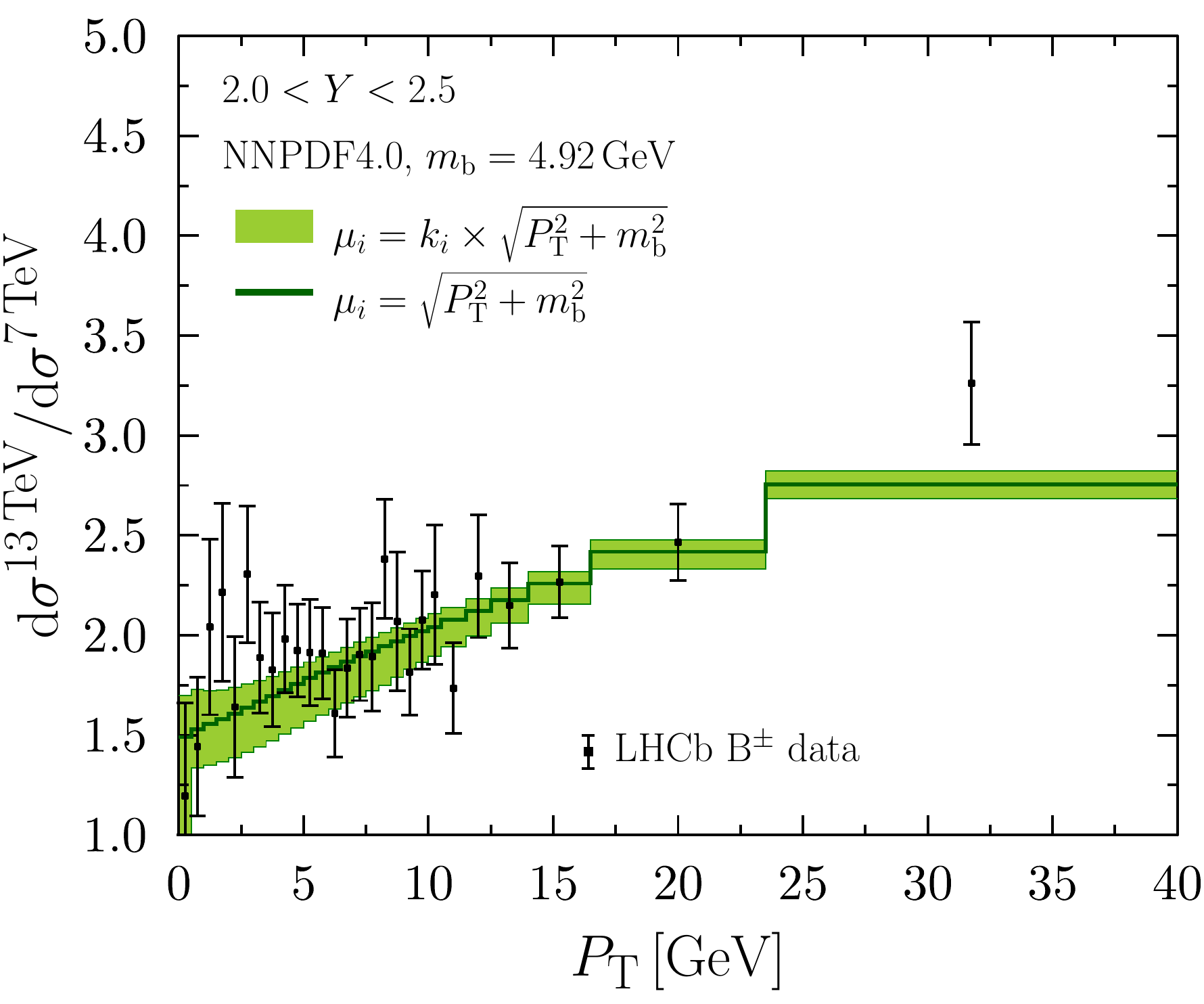}
\includegraphics[width=0.495\linewidth]{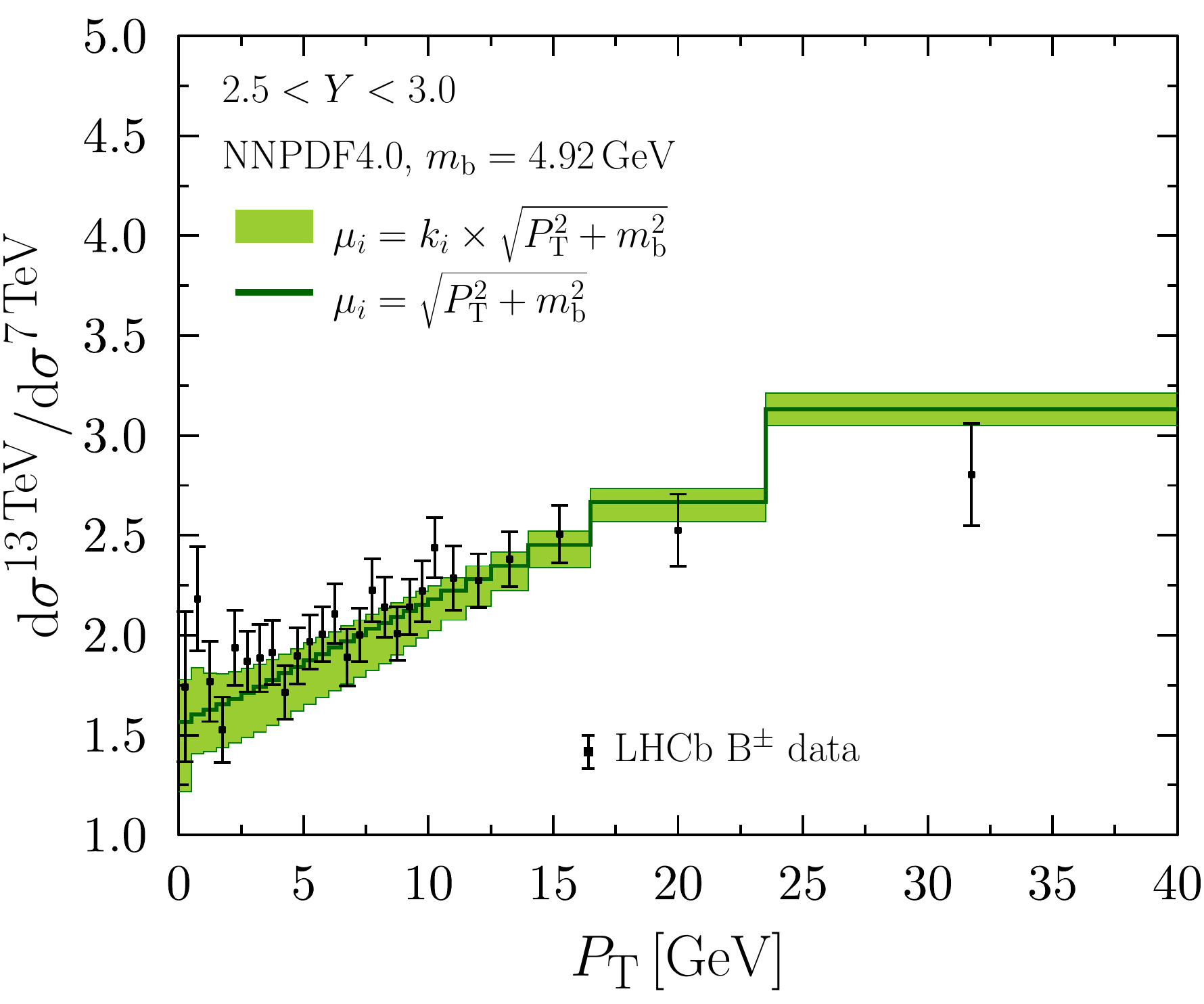}
\includegraphics[width=0.495\linewidth]{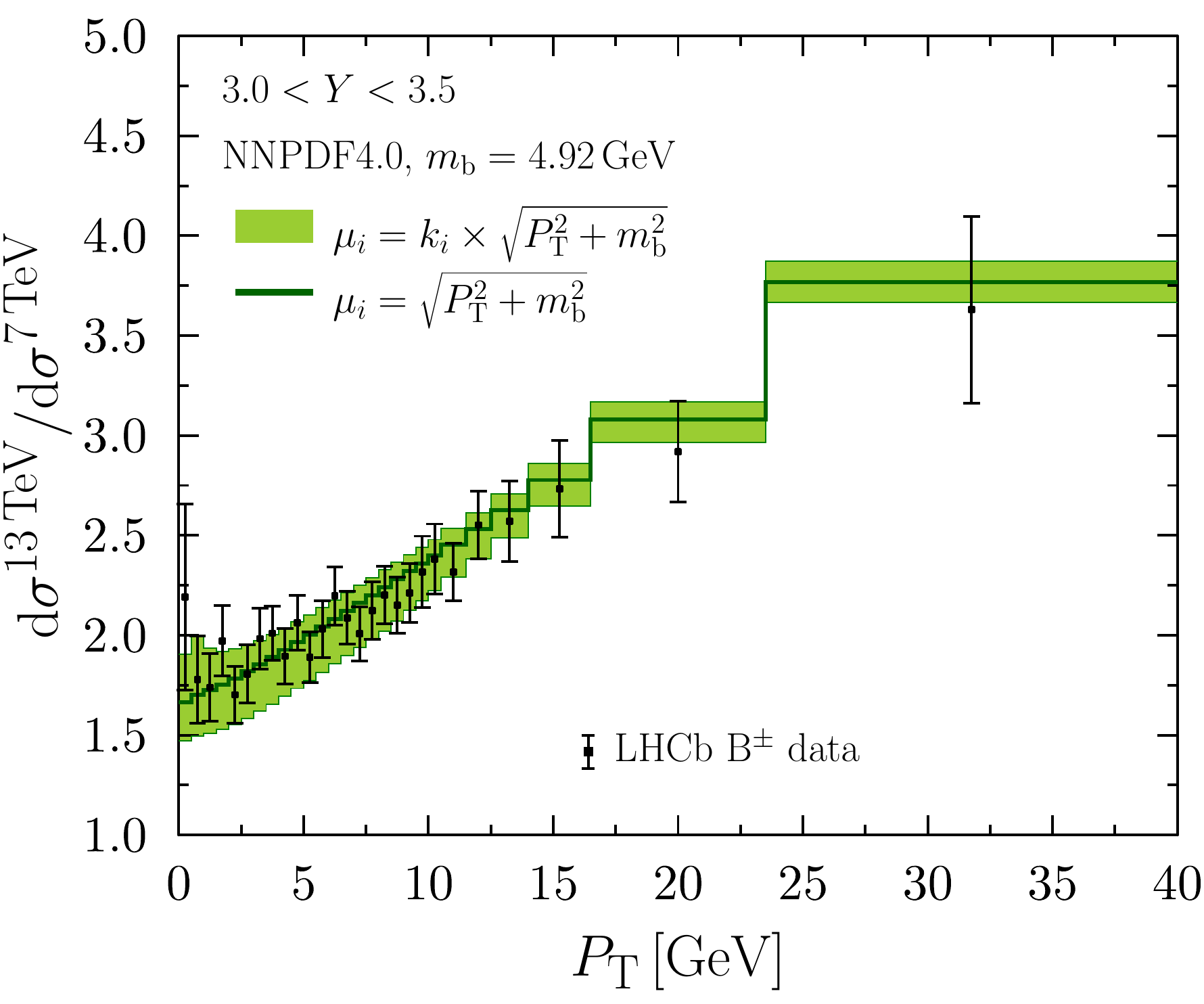}
\includegraphics[width=0.495\linewidth]{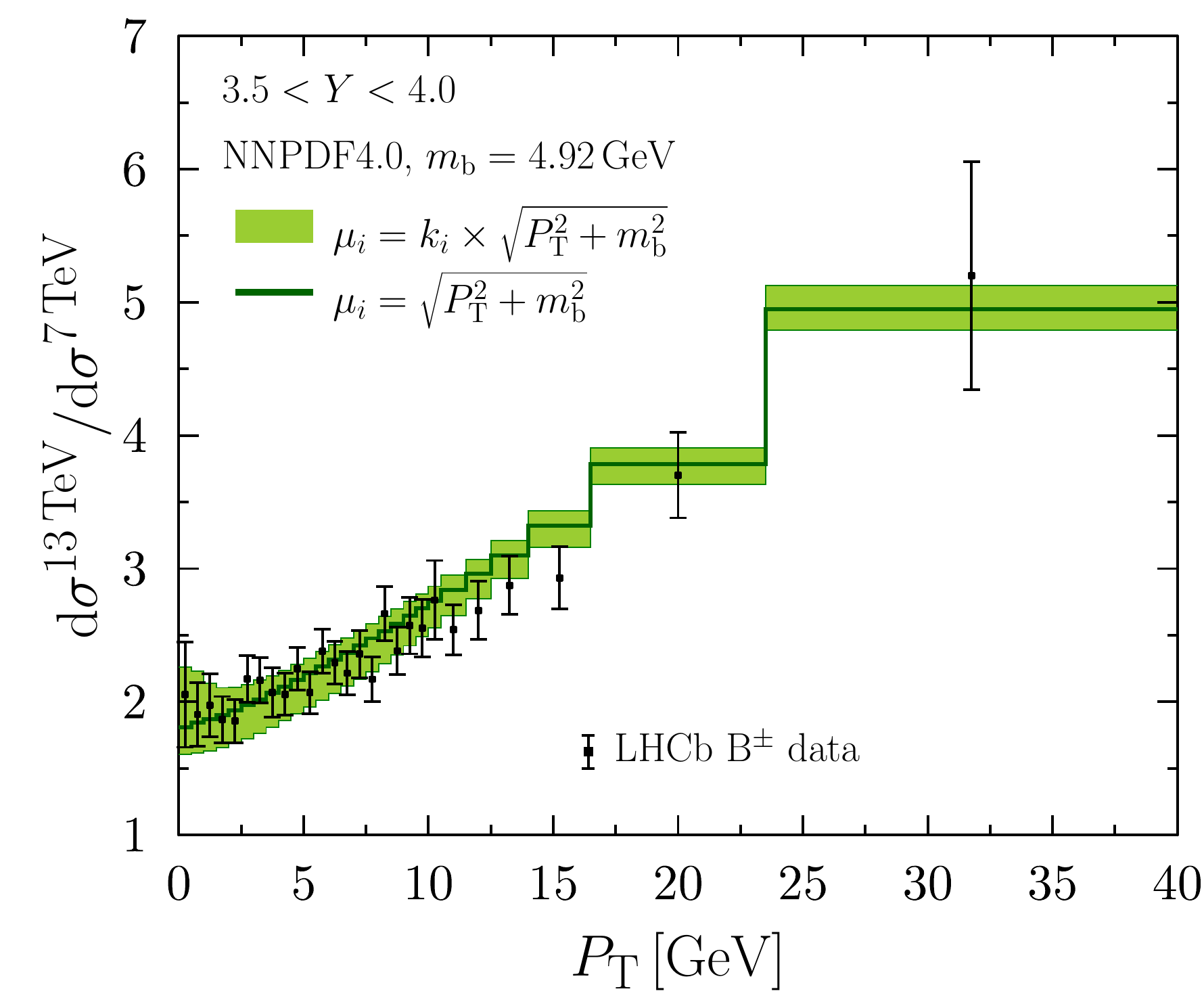}
\includegraphics[width=0.495\linewidth]{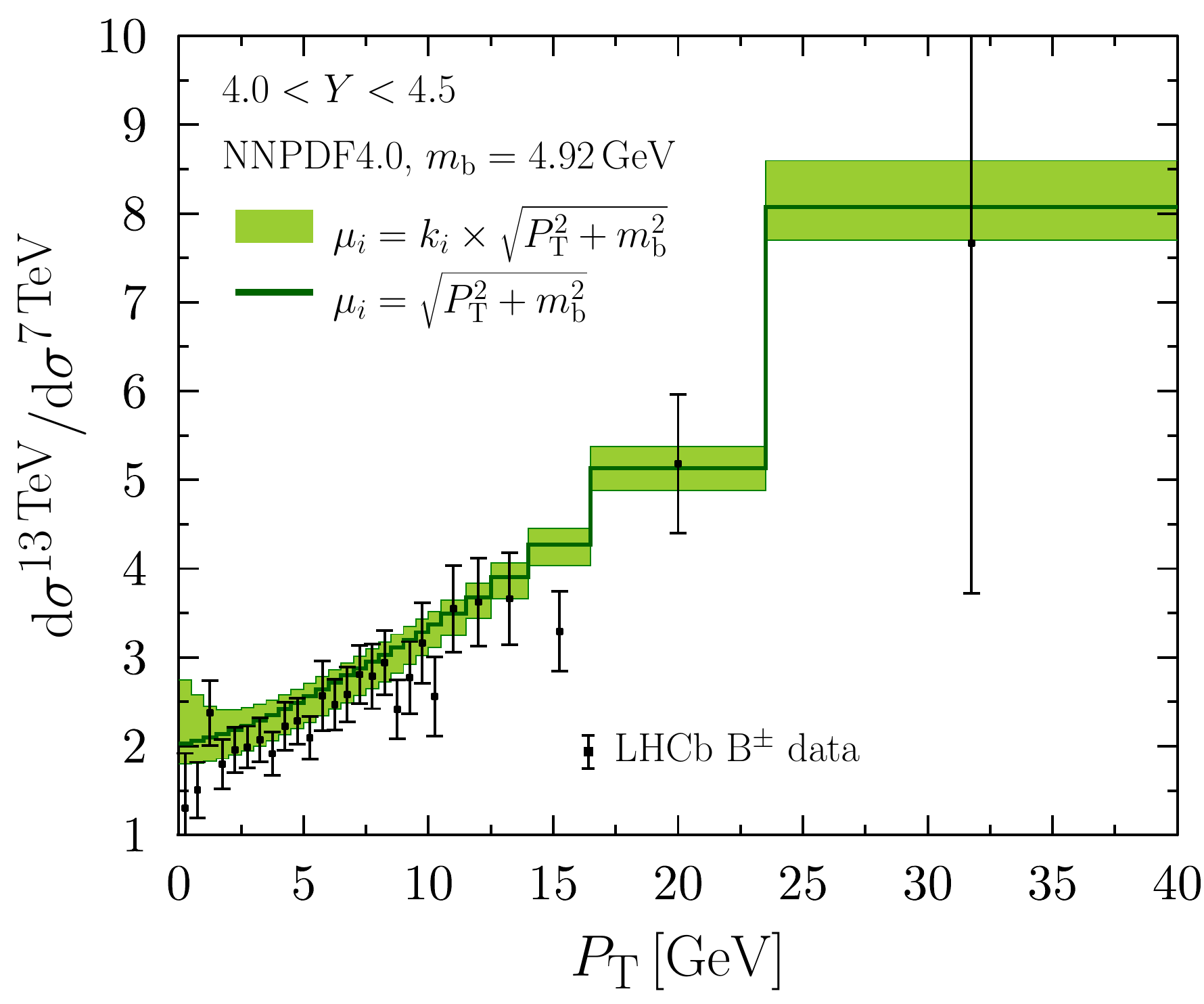}
\caption{
Ratios between the $13\,{\rm TeV}$ and $7\,{\rm TeV}$ B$^\pm$-meson data of the LHCb collaboration \cite{LHCb:2017vec} compared with the SACOT-$m_{\rm T}$ calculation. Each panel correspond to a different rapidity window. The green solid curves show the results of our central scale choice $\mu_i = \sqrt{P_{\rm T}^2 + m_{\mathrm{b}}^2}$, and the light-green filled bands correspond to the uncertainty due to the scale choice.
}
\label{fig:LhCb13vs7TeV}
\end{figure} 

\begin{figure}[htb!]
\centering
\includegraphics[width=0.495\linewidth]{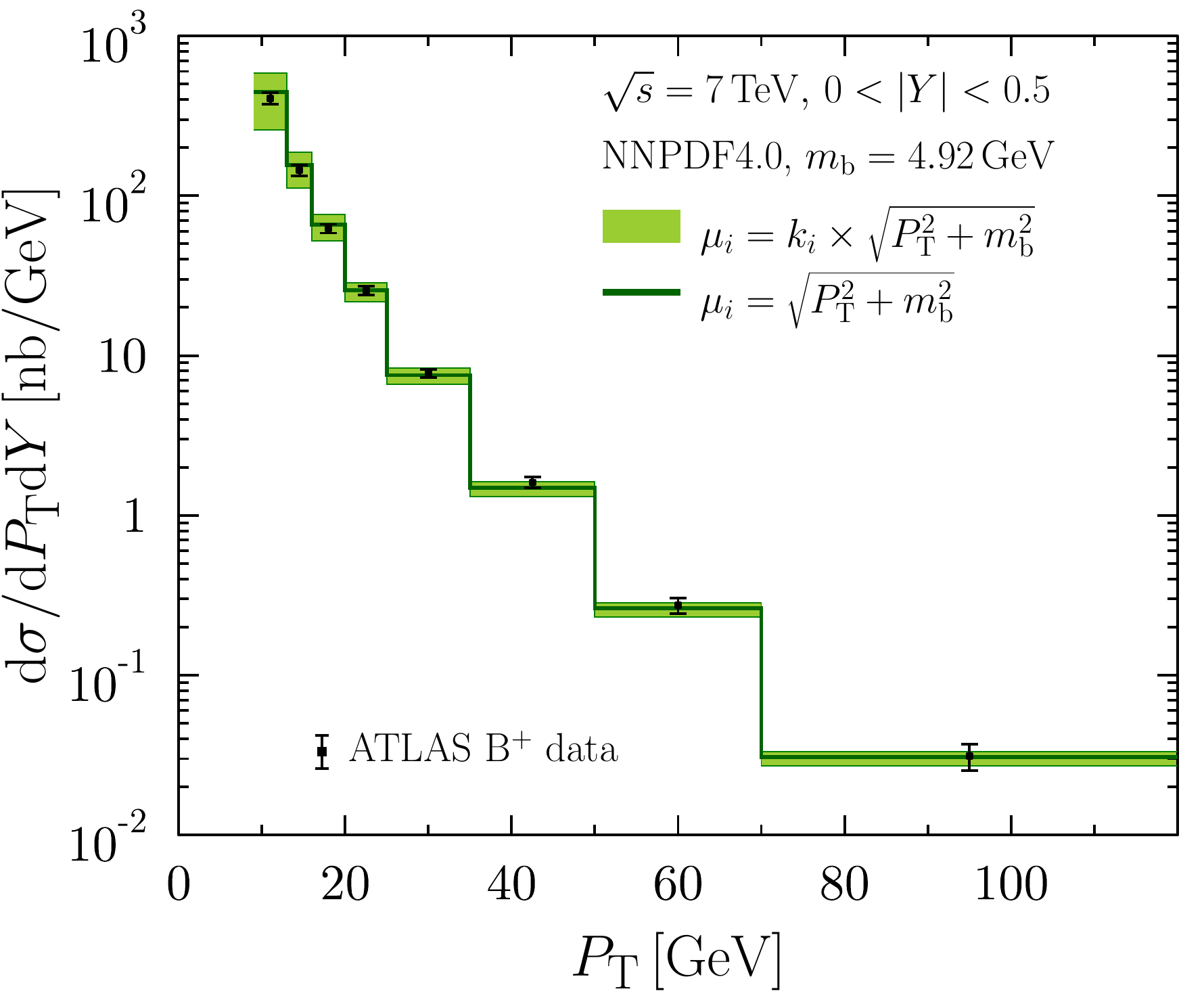}
\includegraphics[width=0.495\linewidth]{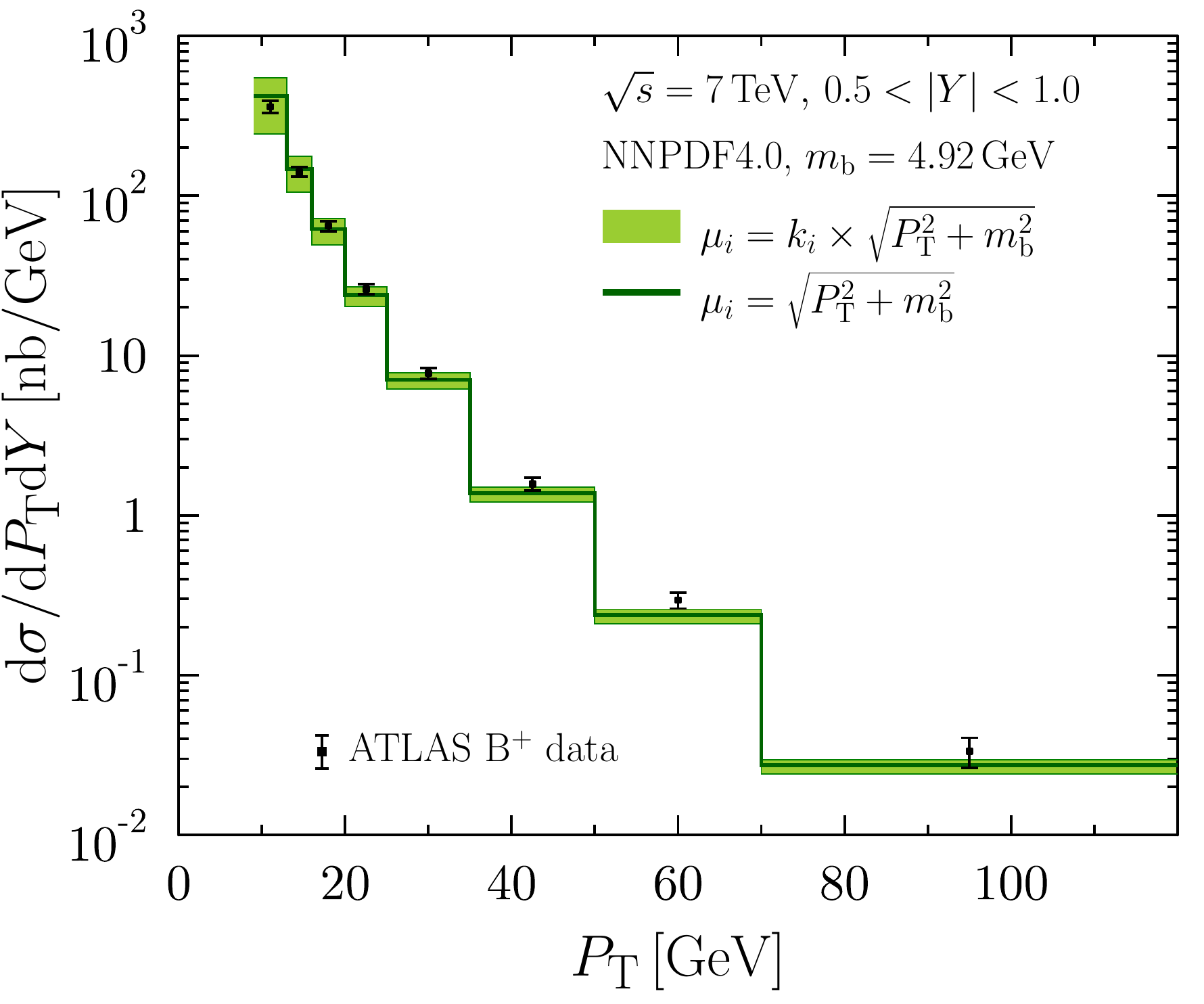}
\includegraphics[width=0.495\linewidth]{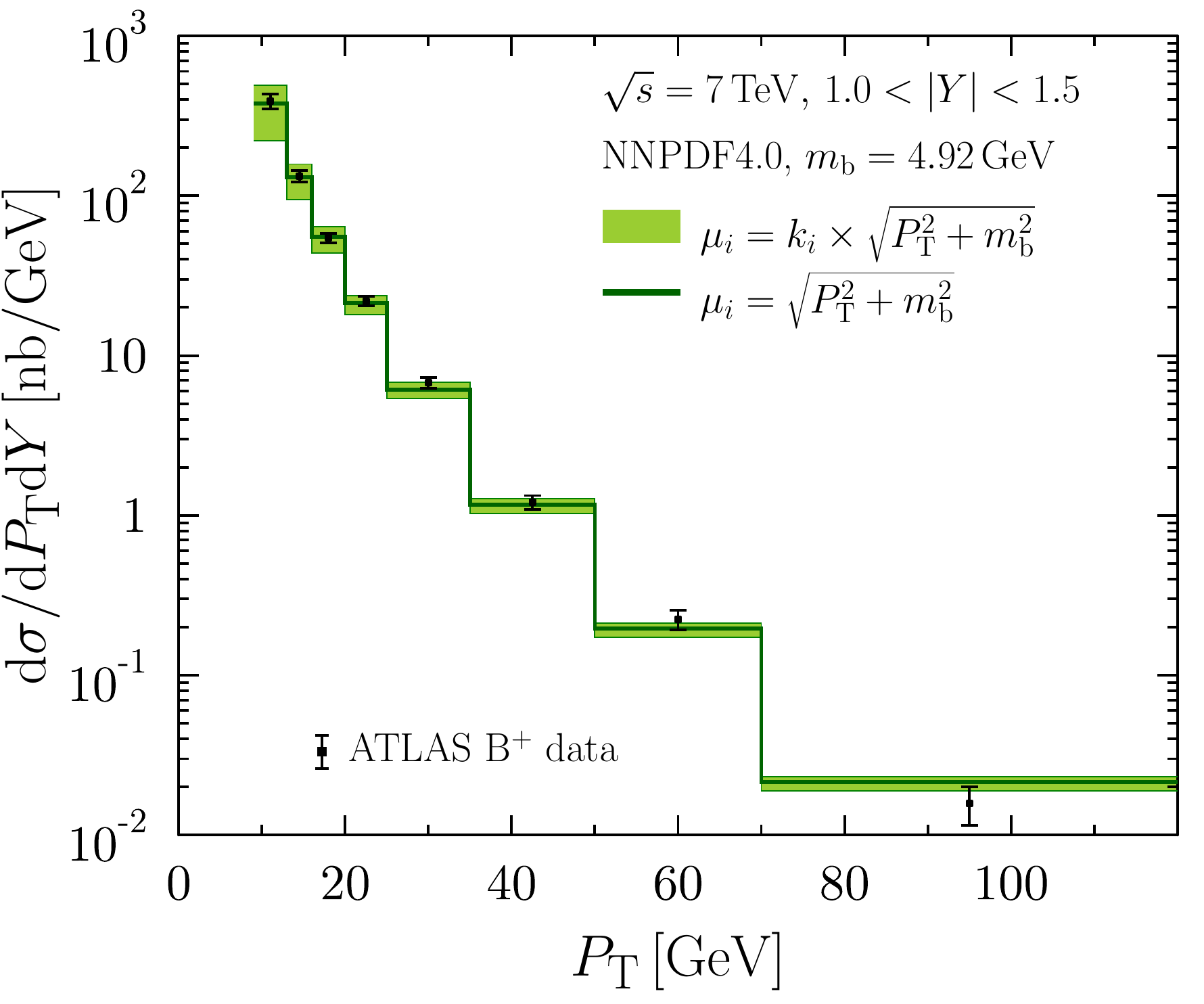}
\includegraphics[width=0.495\linewidth]{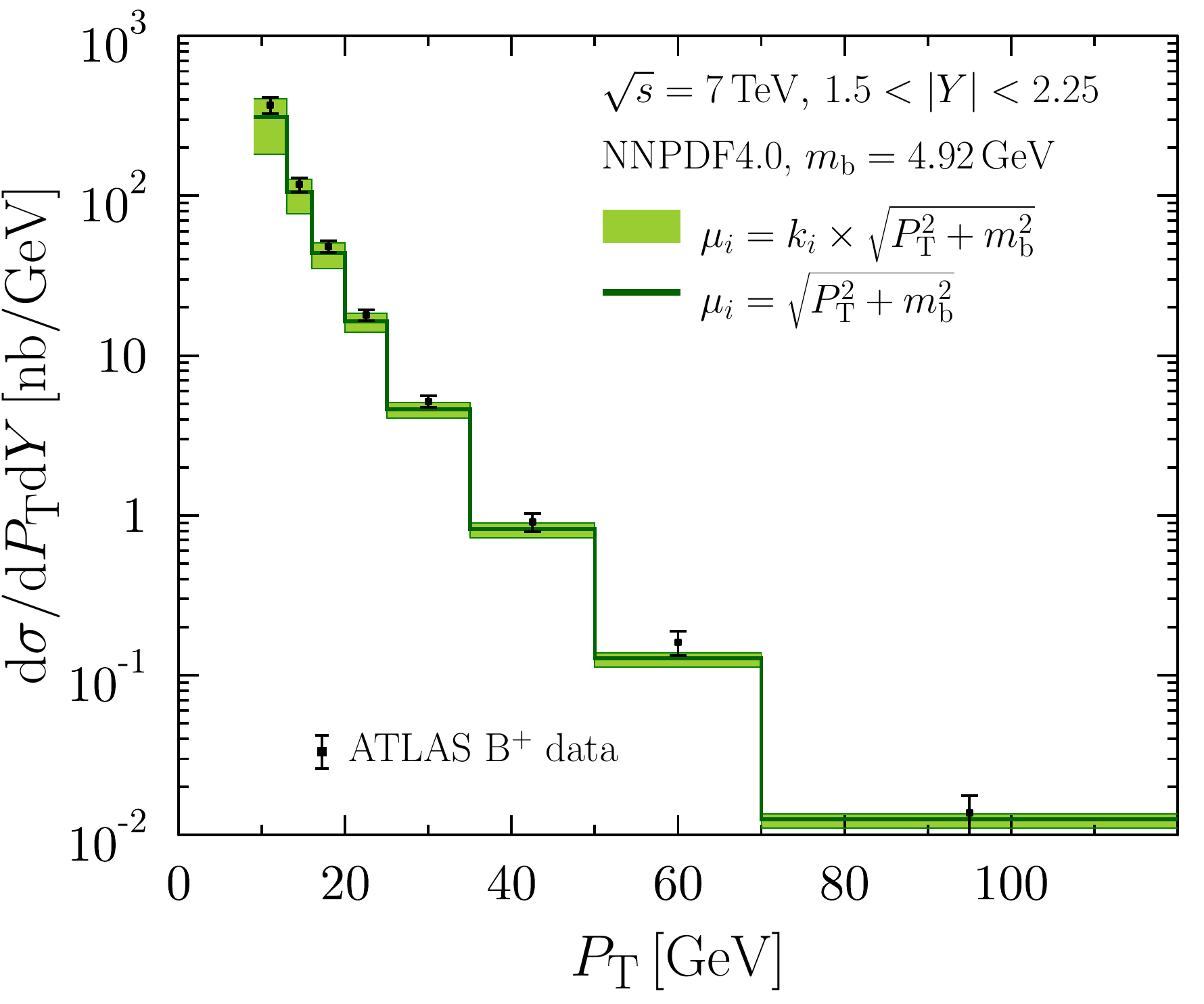}
\includegraphics[width=0.495\linewidth]{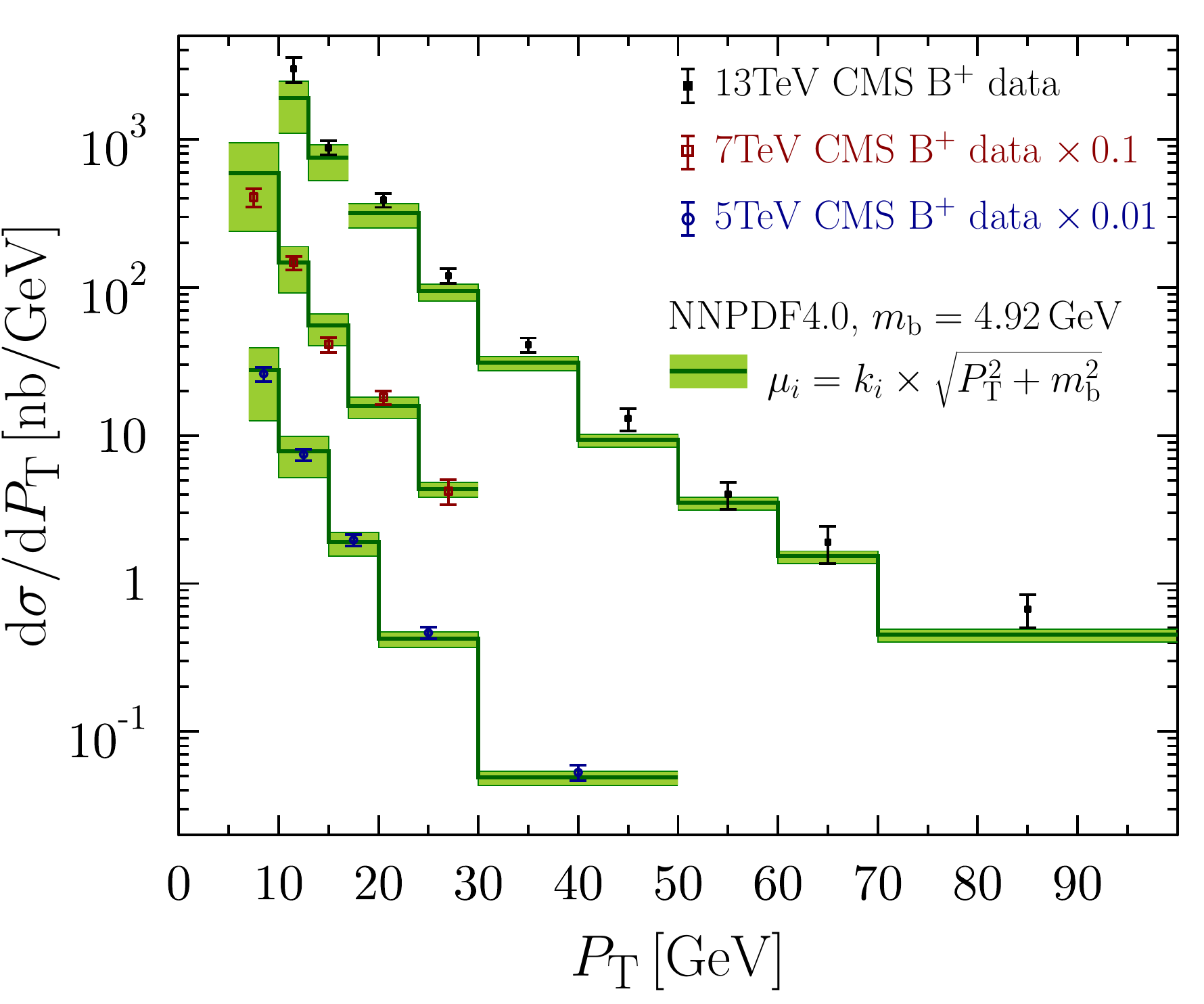} 
\caption{
The $7\,{\rm TeV}$ B$^\pm$-meson data of the ATLAS collaboration \cite{ATLAS:2013cia} (upper panels), and the $5\ldots13\,{\rm TeV}$ B$^\pm$-meson data of the CMS collaboration \cite{CMS:2011oft,CMS:2017uoy,CMS:2016plw} (lowest panel) compared with the SACOT-$m_{\rm T}$ calculation. Each panel correspond to a different rapidity window. The green solid curves show the results of our central scale choice $\mu_i = \sqrt{P_{\rm T}^2 + m_{\mathrm{b}}^2}$, and the light-green filled bands correspond to the uncertainty due to the scale choice.
}
\label{fig:ATLAS_CMS7TeV}
\end{figure} 

The dependence of our calculations on the adopted set of PDFs with different bottom-quark masses is illustrated in Figure~\ref{fig:differentPDFs}. Instead of NNPDF4.0, we have used here the \textsc{MSHT20nlo\_mbrange\_nf5} partons \cite{Cridge:2021qfd}. In this latter analysis, the authors repeated the MSHT20 \cite{Bailey:2020ooq} global PDF fit seven times varying the bottom-quark mass in the range $m_{\mathrm{b}} = 4.0\ldots5.50 \,{\rm GeV}$. By using these PDF sets, we can thus study the bottom-quark dependence of our calculation with the proper behaviour of PDFs (i.e. vanishing bottom-quark) at the threshold $\mu_{\rm fact} = m_{\mathrm{b}}$. The largest differences appear at low $P_{\rm T}$ where the bottom-quark thus plays the most significant role. We see that adopting a smaller bottom-quark mass leads to an increased cross section as the ``mass screening'' in the propagators decreases. Decreasing the bottom-quark mass can be also seen to slightly shift the maximum of the $P_{\rm T}$ spectrum towards lower values of $P_{\rm T}$. The mass dependence is still clearly inferior to the scale dependence of our results i.e. within the scale uncertainties all the shown PDFs agree with the LHCb data (note, however, that variations in the mass also affect the scale choices).

Figures \ref{fig:LhCb7TeV} and \ref{fig:LhCb13TeV} show the comparisons with the LHCb $7\,{\rm TeV}$ and $13\,{\rm TeV}$ \cite{LHCb:2017vec} data using NNPDF4.0 PDFs. In both cases the predictions agree very well with the data throughout the wide rapidity range. The uncertainties from NNPDF4.0 are small (not much larger than the line width) in contrast to the scale uncertainties and are therefore not shown. We consider also the cross section ratios between the collision energies of $13\,{\rm TeV}$ and $7\,{\rm TeV}$. The LHCb paper \cite{LHCb:2017vec} does not contain these ratios separately for different rapidity bins, and we have therefore formed the ratios ourselves from the tabulated cross sections. The statistical and systematical uncertainties have been added in quadrature apart from the 3.9\% systematic uncertainty on the $B^\pm \rightarrow J/\psi K^\pm$ branching fraction (the decay mode measured by the LHCb), which has been canceled out. The results are shown in Figure \ref{fig:LhCb13vs7TeV}. The uncertainties due to the scale choices are vastly smaller in these ratios in comparison to the absolute cross sections. The systematics of the data are well reproduced by the calculation. Despite the smaller scale uncertainties, they are still larger than the PDF-originating uncertainties, at least for NNPDF4.0 which we use here. Finally, Figure \ref{fig:ATLAS_CMS7TeV} presents the ATLAS $7\,{\rm TeV}$ data \cite{ATLAS:2013cia} and the CMS midrapidity data at $5\,{\rm TeV}$ \cite{CMS:2017uoy}, $7\,{\rm TeV}$ \cite{CMS:2011oft}, and $13\,{\rm TeV}$ \cite{CMS:2016plw}. These data do not reach to the low-$P_{\rm T}$ region where most of the bottom-quark mass effects reside but instead extend to higher values of $P_{\mathrm{T}}$ and provide therefore a complementary validation of our computational setup. The dependence of experimental cross sections on the c.m. energy, $P_{\rm T}$ and rapidity are again well reproduced by the calculation.

\begin{figure}[htb!]
\centering
\includegraphics[width=0.8\linewidth]{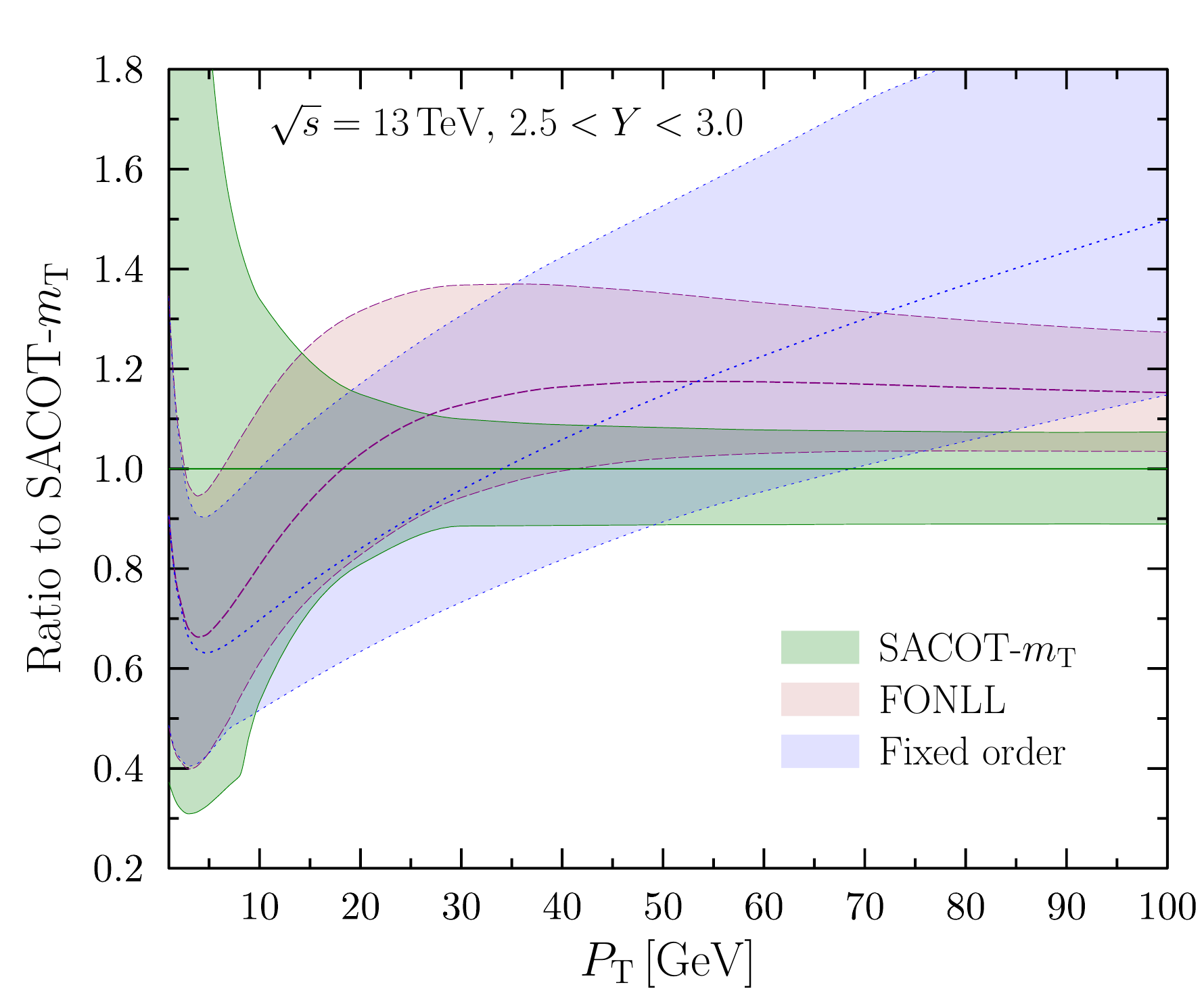}
\caption{
Ratios of FONLL (red band) and fixed-order NLO (blue band) cross sections with respect to the SACOT-$m_{\rm T}$ calculation (green band) in the rapidity window $2.5 < Y < 3.0$ at $\sqrt{s} = 13\,{\rm TeV}$. The FONLL and fixed-order results use the NNPDF3.0 proton PDFs. The widths of the bands correspond to the scale uncertainties of each calculation. 
}   
\label{fig:FONLLvsSACOT}
\end{figure} 

Finally, we wish to illustrate the differences between our SACOT-$m_{\rm T}$ scheme and other approaches. To this end, Figure~\ref{fig:FONLLvsSACOT} presents a comparison in which we have divided the FONLL \cite{Cacciari:1998it,Cacciari:2012ny,Cacciari:2015fta} and fixed-order NLO calculations with the SACOT-$m_{\rm T}$ predictions. The FONLL and fixed-order NLO predictions have been taken from the web interface in Ref.~\cite{FONLL:web} selecting the NNPDF3.0 proton PDFs \cite{NNPDF:2014otw}. The coloured bands show the uncertainties due to the scale variations which, in the case of FONLL and fixer-order calculation, include only variations of the factorization and renormalization scales with 5 different combinations. The FONLL cross section for heavy-quark production is, schematically, of the form,
\begin{align}
d\sigma_q^{\rm FONLL} = \sigma^{\rm fixed \ order} + \frac{p^2_{\rm T}}{p^2_{\rm T} + c^2m_q^2} \big( d\sigma^{\rm resummed} - {\rm subtractions} \big) \,, 
\end{align}
where the default choice $c=5$ has been applied, and which is still folded with a scale-independent fragmentation function to obtain the spectrum of heavy-flavoured mesons. While the fixed-order part includes only those contributions in which the heavy quarks are explicitly produced, the resummed part performs the same resummation of collinear logarithms as the SACOT-$m_{\rm T}$ scheme. The subtraction terms ensure that no double counting takes place. The principal difference with respect to the SACOT-$m_{\rm T}$ scheme is that the resummed part uses pure zero-mass coefficient functions which diverge towards zero $p_{\rm T}$. The factor ${p^2_{\rm T}}/(p^2_{\rm T} + c^2m_q^2)$ is there to tame the divergence. The constant $c$ controls how quickly the resummation is allowed to kick in as a function of $p_{\rm T}$. The default FONLL predictions, however, do not involve an uncertainty due to the variations of the constant $c$. At low $P_{\rm T}$ the FONLL predictions match with the fixed-order calculations and show a clearly smaller scale uncertainty in comparison to the SACOT-$m_{\rm T}$ scheme. This is due to the fact that FONLL suppresses the contributions from the resummed part (which comes with a large scale uncertainty at low $P_{\rm T}$) by choosing a large enough $c$. This is of course well justified in the sense that at low $P_{\rm T}$ the collinear logarithms are not yet large and thus their resummation cannot be a big effect either. However, we recall that by including the $\mathcal{O}(\alpha_s^3)$ terms in the resummed cross sections, they also effectively contain contributions from the fixed-order NNLO calculations which can be significant even if the resummation of the associated logarithms is not yet crucial. Moving towards somewhat higher values of $P_{\rm T}$ the scale uncertainty of the SACOT-$m_{\rm T}$ scheme quickly diminishes and becomes eventually smaller than that of FONLL -- starting from $P_{\rm T} \approx 3m_{\rm b}$ or so. This indicates that the resummation begins to have an effect at such values of $P_{\rm T}$ but the chosen value of $c$ in FONLL still retains the fixed-order contribution (with a larger scale uncertainty) significant. At the high-$P_{\rm T}$ end both FONLL and SACOT-$m_{\rm T}$ display a scale uncertainty which is approximately the same for both and clearly smaller than the scale uncertainty of the fixed-order predictions.  

\section{Results for proton-nucleus collisions}
\label{sec:results-pPb}

The D-meson production in p-Pb collisions \cite{LHCb:2017yua} has been used as a constraint in the EPPS21 \cite{Eskola:2021nhw} and nNNPDF3.0 \cite{Khalek:2022zqe} fits of nuclear PDFs. The theoretical framework in the EPPS21 analysis was the one discussed here, SACOT-$m_{\rm T}$, while the nNNPDF3.0 analysis used a fixed-order POWHEG calculation \cite{Nason:2004rx,Frixione:2007vw,Alioli:2010xd} matched to the PYTHIA \cite{Bierlich:2022pfr} parton shower. The differences between the two approaches were discussed in Ref.~\cite{Helenius:2018uul}. Heavy-flavour observables have also been studied in a recent variant of the nCTEQ15 analysis \cite{Duwentaster:2022kpv} but using considerable simplifications on the partonic matrix elements and kinematics. Specifically, it is the nuclear modification 
\begin{align}
R_{\rm pPb} = \frac{d^2\sigma^{\rm p\text{-}Pb}/dYdP_{\rm T}}{d^2\sigma^{\rm p\text{-}p}/dYdP_{\rm T}}  \,,
\end{align}
for D-meson production that enters the EPPS21 and nNNPDF3.0 analyses. In such ratio most of the scale uncertainties in the SACOT-$m_{\rm T}$ scheme were observed to cancel between the numerator and denominator though some dependence persist, particularly at low $P_{\rm T}$ \cite{Eskola:2019bgf}. To be on the safe side, EPPS21 imposed a cut $P_{\rm T} > 3 \, {\rm GeV}$. In the POWHEG+PYTHIA approach the scale uncertainties in $R_{\rm pPb}$ at high $P_{\rm T}$ were observed to be much larger than in the SACOT-$m_{\rm T}$ scheme \cite{Khalek:2022zqe}. The nNNPDF3.0 analysis nevertheless included the D-meson data without any restrictions in $P_{\rm T}$ excluding, however, the p-Pb data at backward rapidities ($Y < 0$). In both cases, the inclusion of the LHCb data \cite{LHCb:2017yua} led to a significant reduction of the nuclear-PDF uncertainties at small $x$. In this section we will now use these D-meson-constrained nuclear PDFs to predict the nuclear modification ratios for B-mesons and see whether the predictions agree with the recent LHCb data \cite{LHCb:2019avm}. 

\begin{figure}[htb!]
\centering
\includegraphics[width=0.495\linewidth]{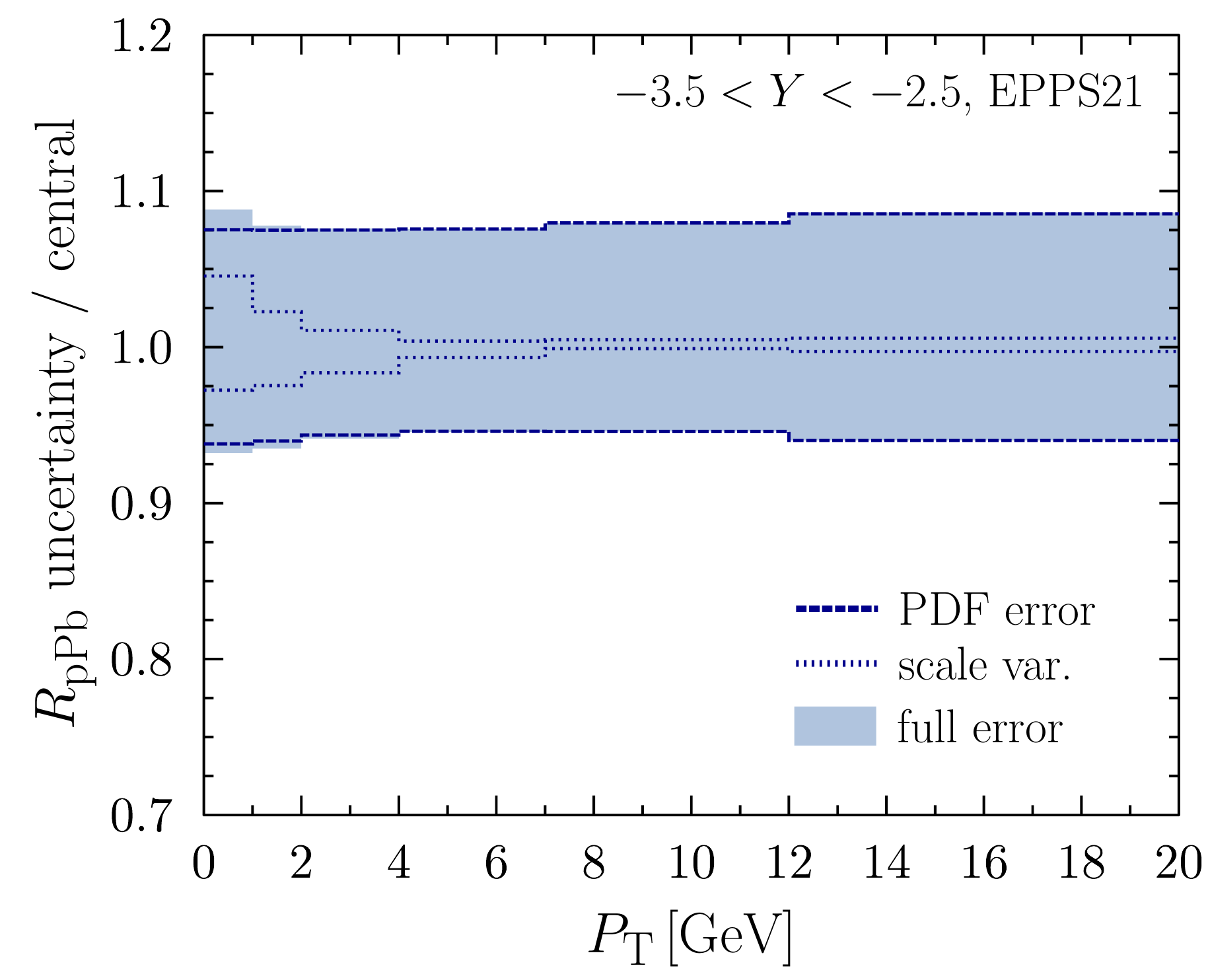}
\includegraphics[width=0.495\linewidth]{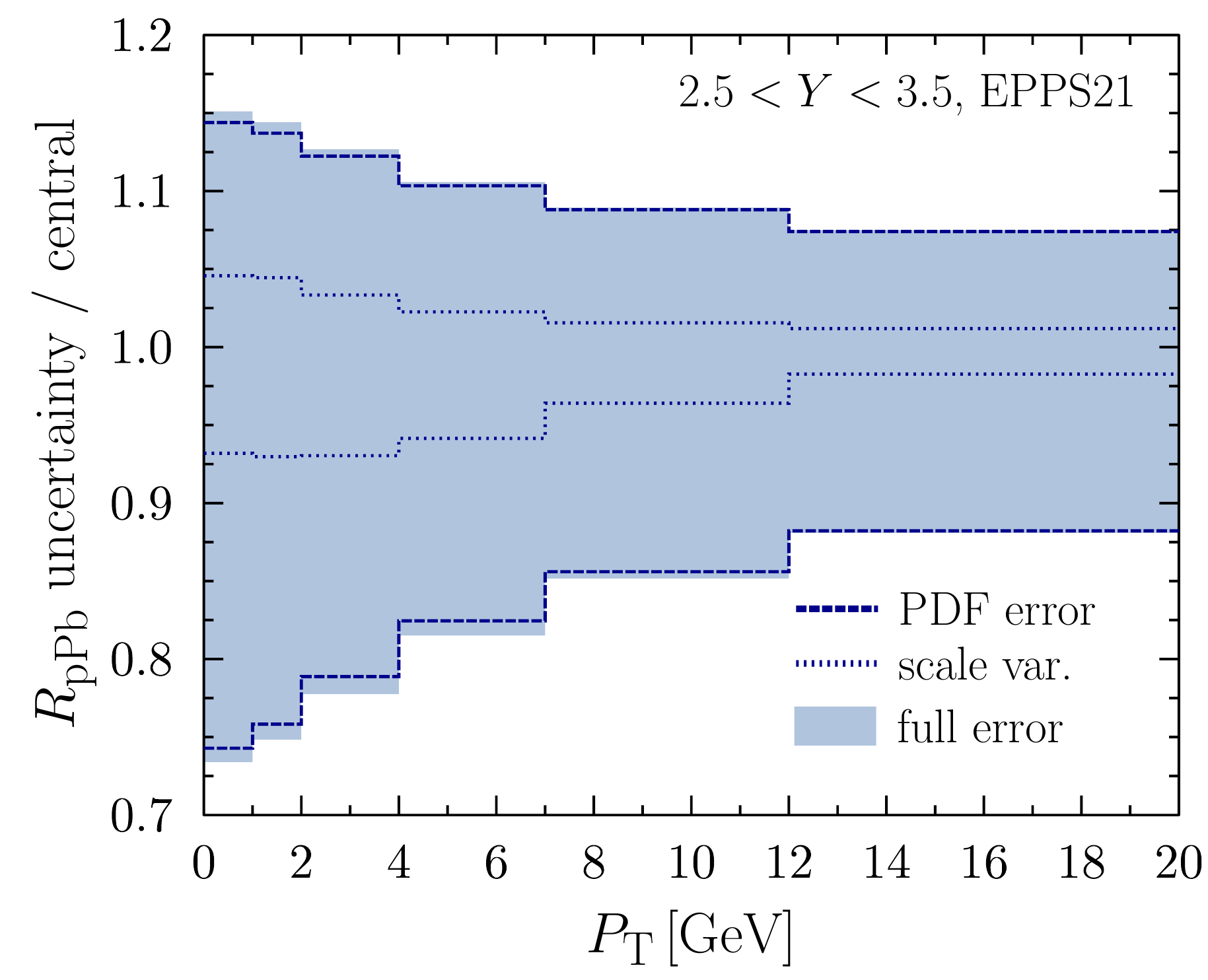}
\includegraphics[width=0.495\linewidth]{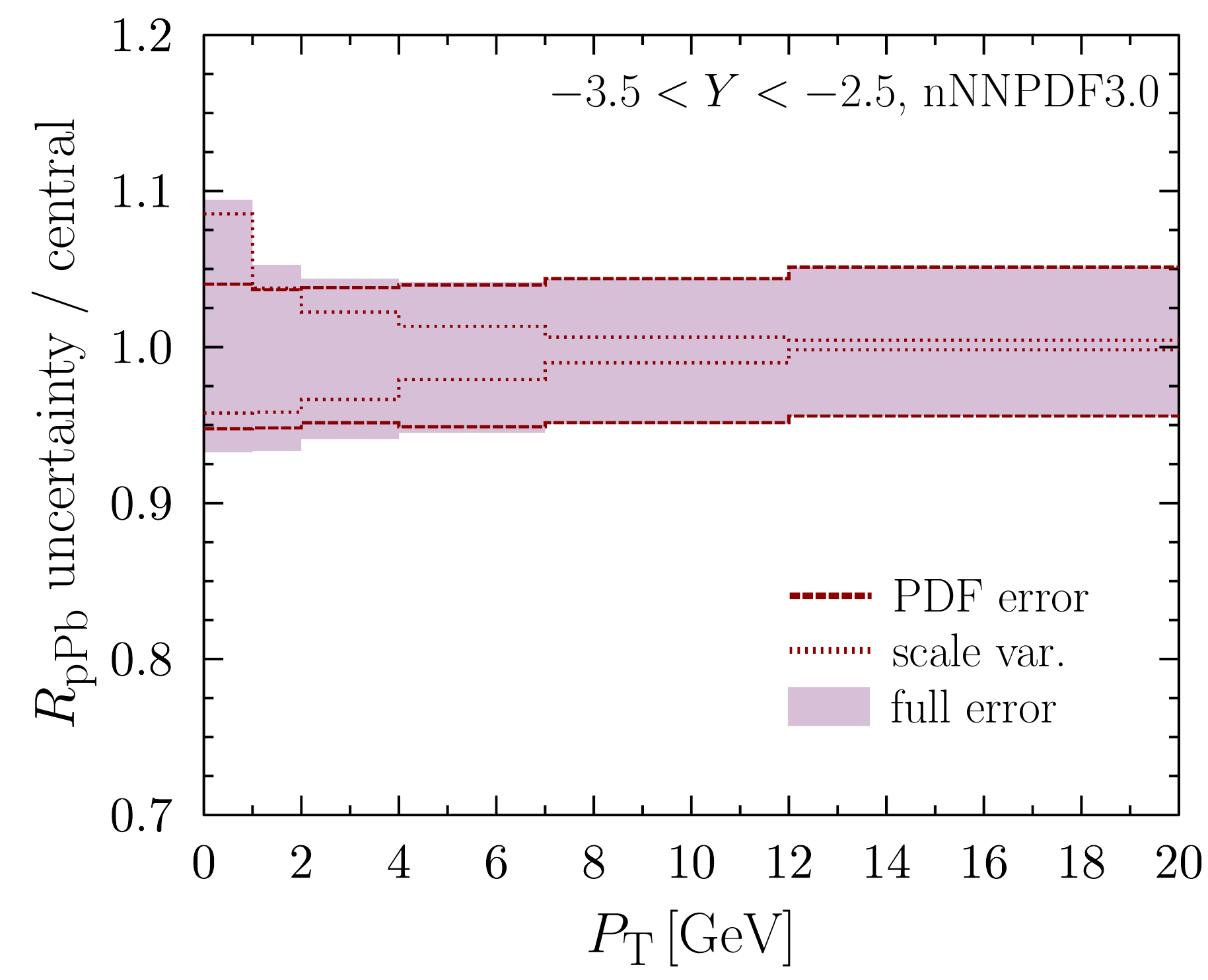}
\includegraphics[width=0.495\linewidth]{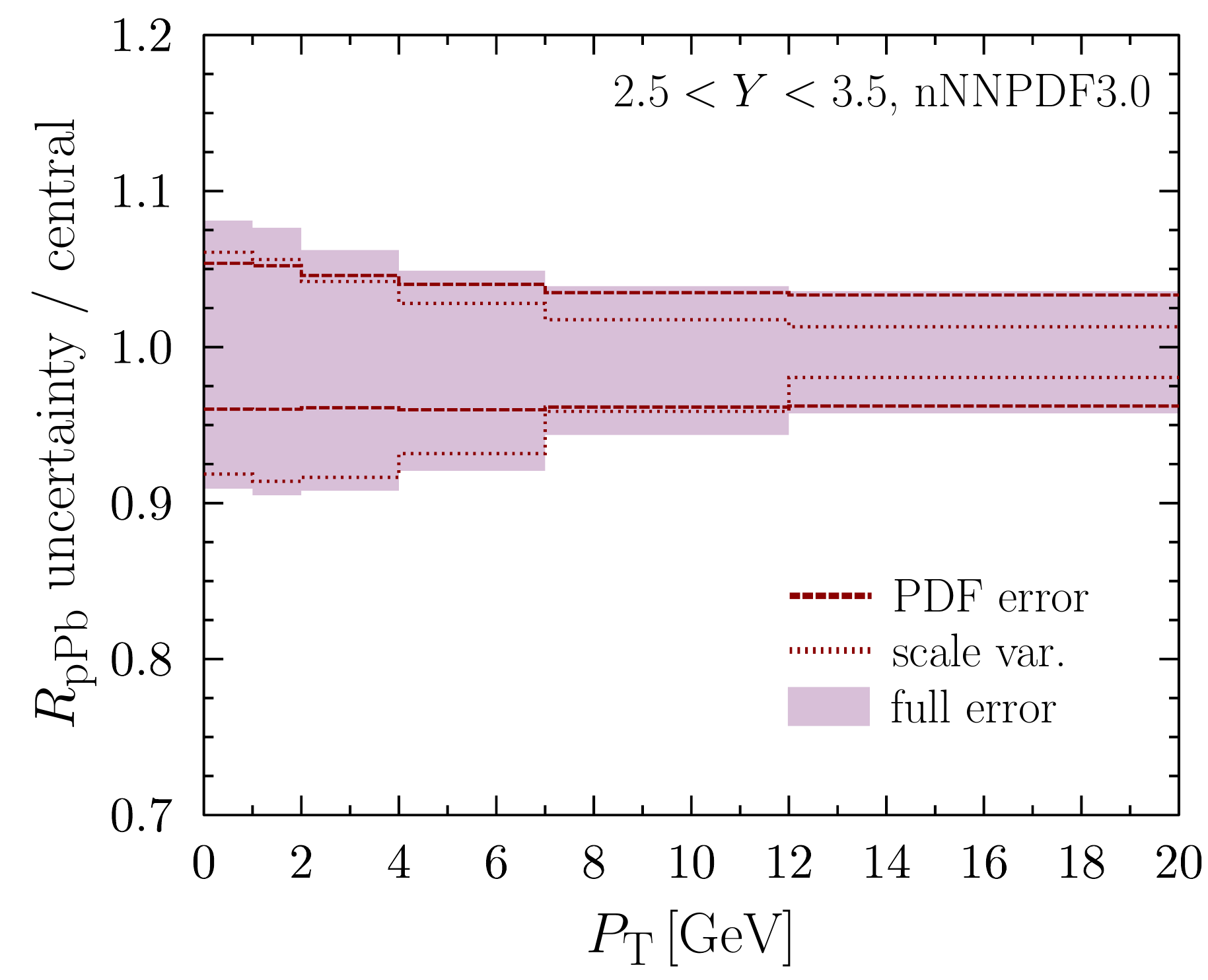}
\caption{
The scale (dotted) and 90\% PDF (dashed) uncertainties of the B-meson nuclear modification factors in p-Pb collisions. The upper panels correspond to the EPPS21 PDFs \cite{Eskola:2021nhw} and the lower panels to the nNNPDF3.0 \cite{Khalek:2022zqe} PDFs. 
}
\label{fig:RpAerr}
\end{figure} 

\begin{figure}[htb!]
\centering
\includegraphics[width=0.495\linewidth]{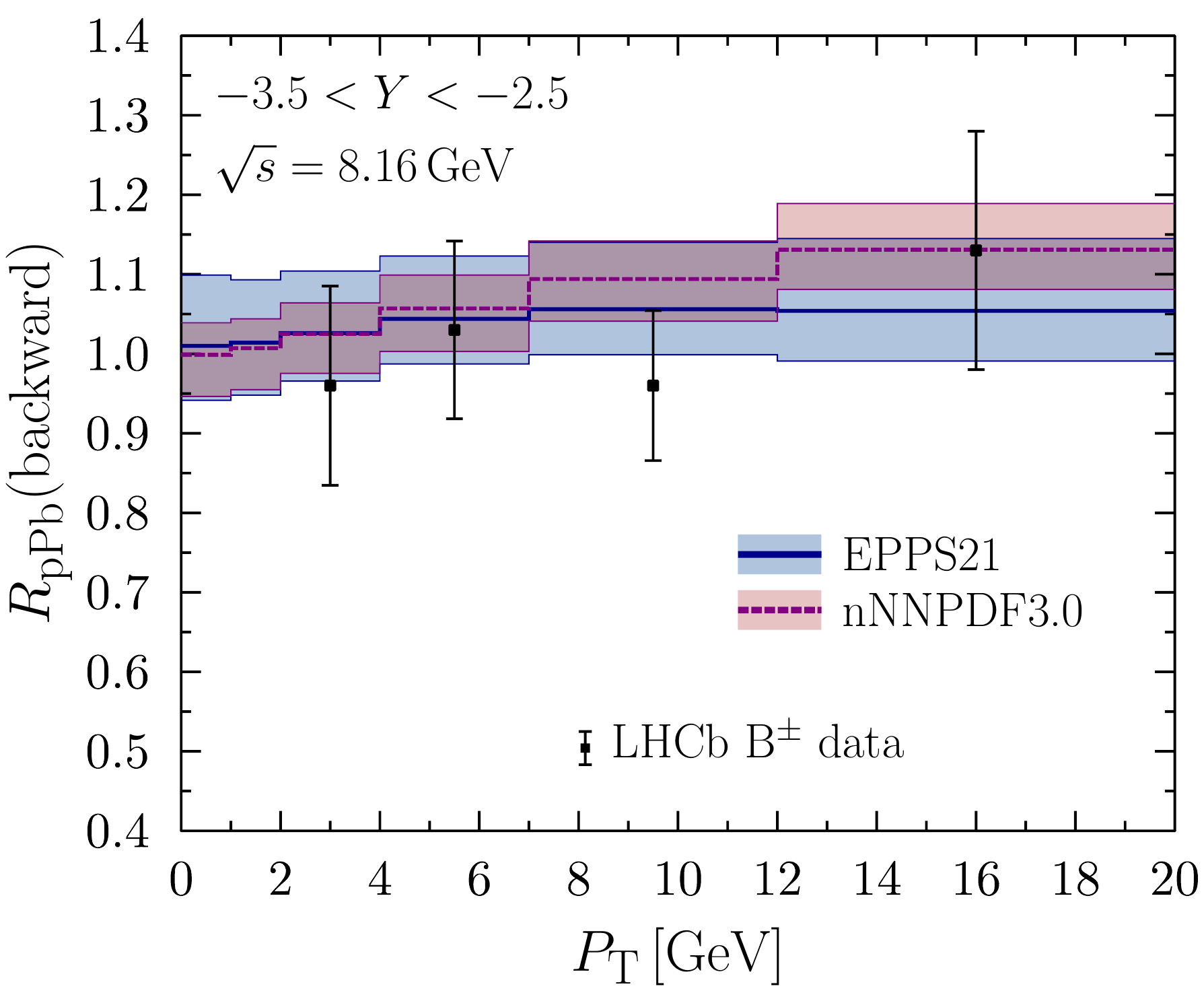}
\includegraphics[width=0.495\linewidth]{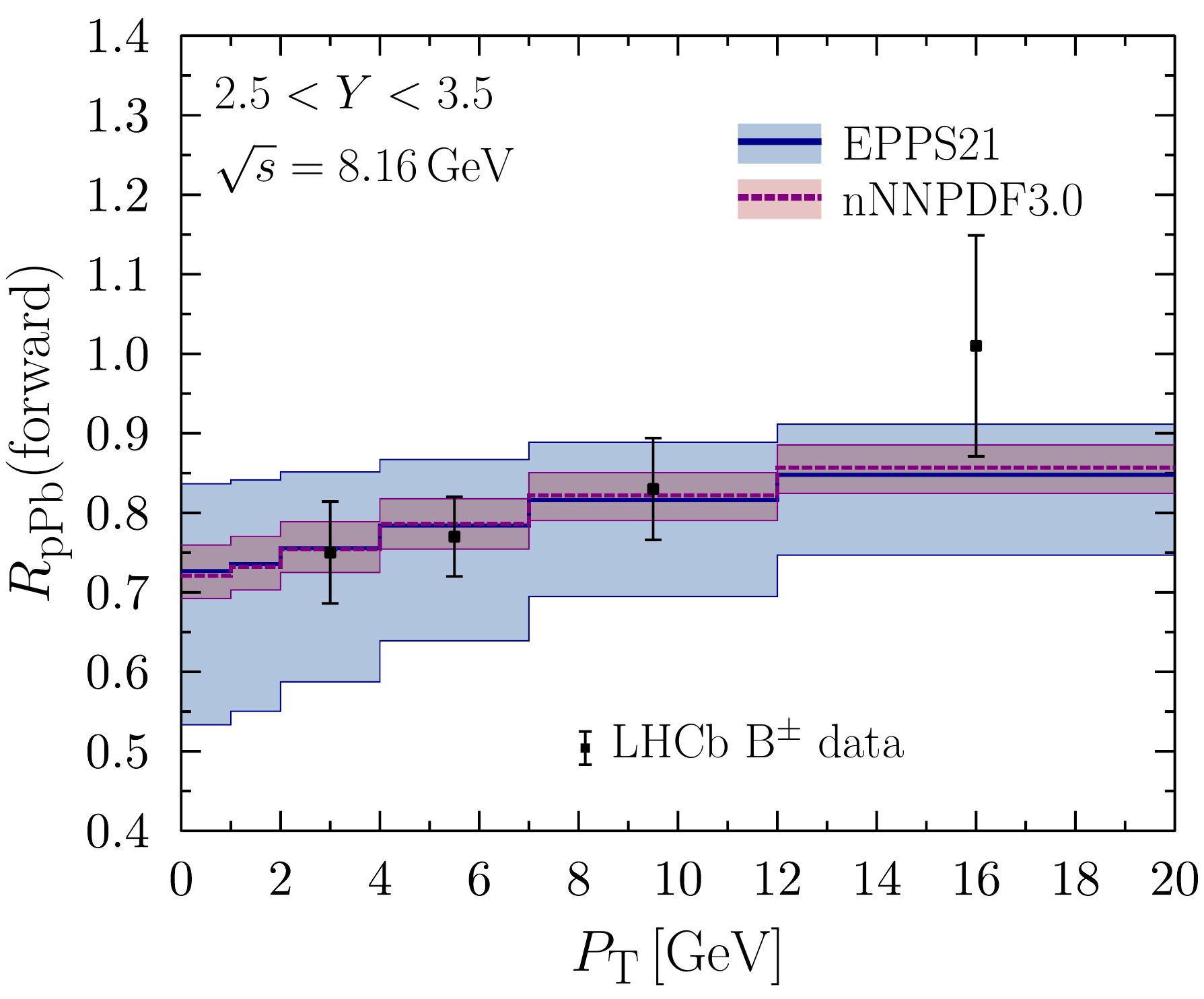}
\includegraphics[width=0.495\linewidth]{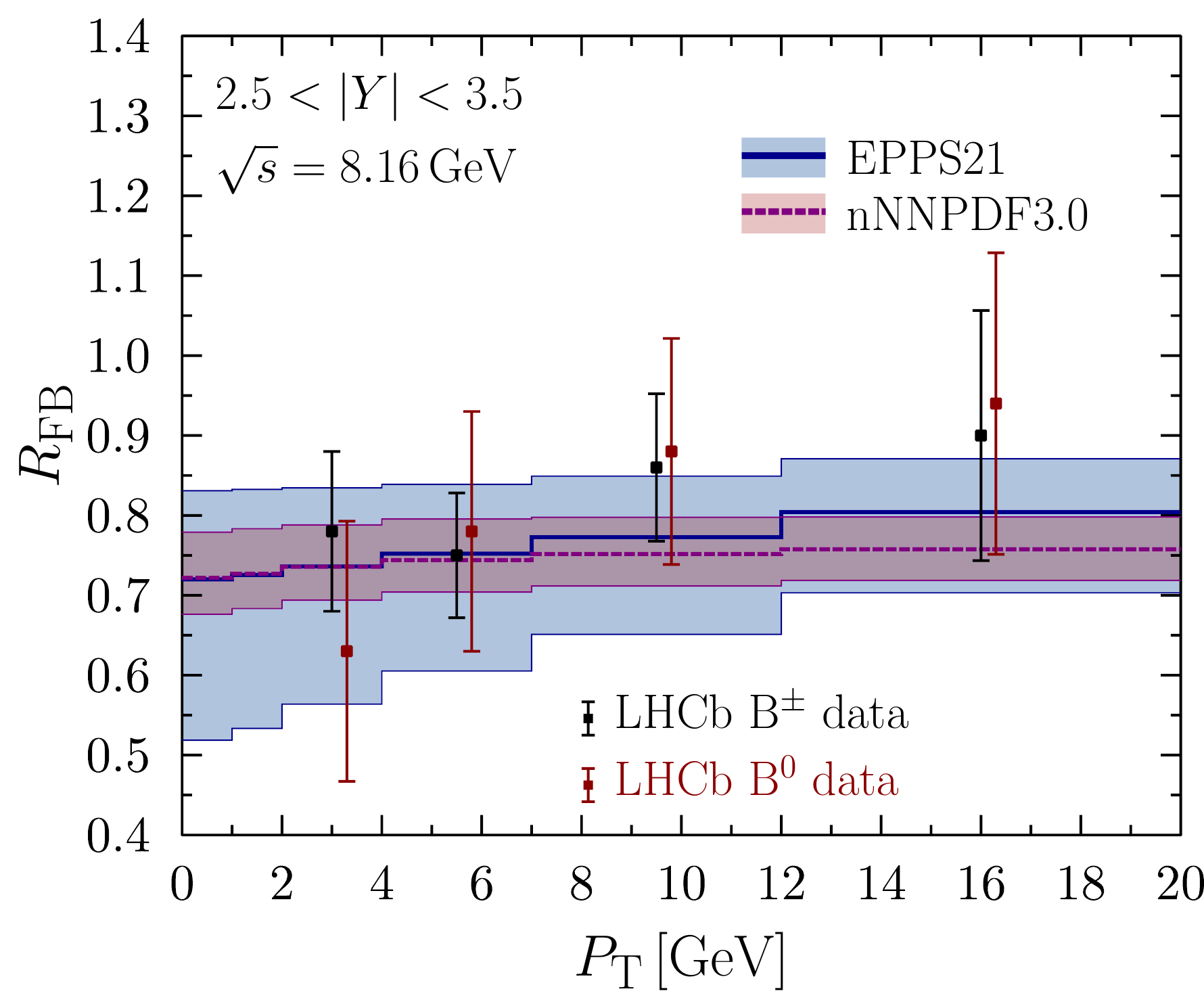}
\caption{The nuclear modification factors (upper panels) and the forward-to-backward ratio (lower panel) for B mesons. The coloured bands correspond to the EPPS21 \cite{Eskola:2021nhw} (blue) and nNNPDF3.0 \cite{Khalek:2022zqe} (purple) nuclear-PDF uncertainties. The data are from Ref.~\cite{LHCb:2019avm}. 
}
\label{fig:RpA}
\end{figure} 

Before comparing with the data we study the relative size of the PDF and scale uncertainties in B-meson $R_{\rm pPb}$. This is done in Figure~\ref{fig:RpAerr} in which the relative scale and 90\% PDF uncertainties for $R_{\rm pPb}$ are shown. For EPPS21, the PDF uncertainty is calculated according to the Hessian prescription, see Sect.~4.3 of Ref.~\cite{Eskola:2021nhw}, whereas the 90\% nNNPDF3.0 uncertainty is calculated by rejecting the predictions of those replicas that constitute 10\% of the most extreme predictions, see Sect.~7.2 of Ref.~\cite{Khalek:2022zqe}. In both cases, the correlations between the nuclear and proton PDFs are accounted for. The full uncertainty band combines the PDF and scale uncertainties in quadrature. The scale uncertainties are the largest at low values of $P_{\rm T}$ and they are  very similar between EPPS21 and nNNPDF3.0. In the case of EPPS21 the PDF uncertainties are always clearly larger than the those induced by the scale variations. The nNNPDF3.0 PDF uncertainties are, however, systematically smaller than those of EPPS21 and in places the scale uncertainty competes and even exceeds the PDF uncertainty. The fact that the nNNPDF3.0 uncertainties are generally smaller than those of EPPS21 is presumably mostly due to the methodological differences between these two PDF analyses \cite{Courtoy:2022ocu}.

Figure \ref{fig:RpA} shows how our calculations using the EPPS21 and nNNPDF3.0 nuclear PDFs compare against the LHCb B$^\pm$-meson data at $8.16\,{\rm TeV}$ \cite{LHCb:2019avm}. In the backward direction $-3.5 < Y < -2.5$  ($Y$ referring to the rapidity of the meson in nucleon-nucleon center-of-mass frame) one probes predominantly the large-$x$ part of the nuclear PDFs where there is an enhancement (antishadowing) in comparison to the proton PDFs. In the forward direction $2.5 < Y < 3.5$ it is the small-$x$ regime of nuclear PDFs that matters the most where the nuclear PDFs are suppressed (shadowing) in comparison to the proton PDFs. The LHCb data are broadly consistent with these expectations and quantitatively agree with both EPPS21 and nNNPDF3.0. In particular, the data at forward direction are more precise than the EPPS21 predictions and could possibly give some additional constraints in a global analysis of nuclear PDFs, though the statistical weight of these B-meson data will be small in a global $\chi^2$ analysis. The lower panel still shows the forward-to-backward ratio 
\begin{align}
R_{\rm FB} = \frac{d^2\sigma^{\rm p\text{-}Pb}(Y > 0)/dYdP_{\rm T}}{d^2\sigma^{\rm p\text{-}Pb}(Y < 0)/dYdP_{\rm T}}  \,,
\end{align}
which requires no p-p baseline measurement. We also show the B$^0$ measurement -- our calculation is identical for B$^\pm$ and B$^0$ (the KKSS08 FFs are the same for these two species). Note that the data in the forward and backward directions come from separate LHC runs with different beam configurations so the luminosity uncertainties do not cancel. Our predictions are found to agree with the data also here. The data perhaps hints towards a stronger $P_{\rm T}$ dependence but a more precise measurement is still required to confirm this in a statistically significant way as notable fluctuations are seen in LHCb data for $R_{\rm pPb}$ at this $P_{\rm T}$ region.

\section{Conclusion and Outlook}
\label{sec:Conclusion}

In summary, we have extended the NLO SACOT-$m_{\rm T}$ scheme, originally introduced in the context of D-meson production, to the case of B-meson production at the LHC. In the original version we had defined a fragmentation variable that could lead to a pathological behaviour in certain corners of the phase space -- a better version introduced in the present paper evades this problem. We contrasted our calculations against the proton-proton data from the LHCb, ATLAS and CMS collaborations finding a very good agreement within theoretical uncertainties originating from the variations of the renormalization/factorization/fragmentation scales and the bottom-quark mass. Notably, the shift in the position of the peak value in $P_{\mathrm{T}}$ spectra when increasing the heavy-quark mass is naturally reproduced with our default setup. We found a good agreement also with the data at high-$P_{\mathrm{T}}$ region where the scale variations play a smaller role. To get some more insight on different GM-VFNS schemes, we compared our results to the FONLL approach and concluded that the somewhat different evolution of scale uncertainties as a function of $P_{\mathrm{T}}$ can be attributed to a different regulation of massless coefficient functions which also controls the relative contribution of direct and non-direct production channels. While the scale uncertainties can be large in the case of absolute cross sections, they are strongly suppressed e.g. in ratios of cross sections between different c.m. energies or ratios between different collision systems, which are then much more sensitive to the underlying proton and/or nuclear structure. In particular, we considered the nuclear modification $R_{\rm pPb}$ and the forward-to-backward ratio $R_{\mathrm{FB}}$ by using the EPPS21 and nNNPDF3.0 nuclear PDFs. The predictions agree very well with the data from the LHCb collaboration, lending further support for the universality of nuclear PDFs. 

Having now tested the SACOT-$m_{\rm T}$ scheme in the case of inclusive D- and B-meson production, we plan to extend our framework also to include the decays of these open heavy flavours. In many cases the decay particles -- e.g. the J/$\psi$ spectrum from B mesons -- can be measured with a significantly greater accuracy than the fully reconstructed D or B mesons. This would then open e.g. the possibility to include the corresponding $R_{\rm pPb}$ data to the global fits of nuclear PDFs without resorting to simplifying approximations made in other works and provide more constraints for small-$x$ gluon shadowing in heavy nuclei. In addition, now that the fixed-order NNLO calculations for $b \overline{b}$ production are/will soon be publicly available, it begins to be possible to increase the accuracy of GM-VFNS in hadroproduction to include higher-order perturbative contributions.

\section*{Acknowledgements}

Our work has been supported by the Academy of Finland, projects 308301 and 331545, and was funded as a part of the Center of Excellence in Quark Matter of the Academy of Finland, project 346326. The reported work is associated with the European Research Council project ERC-2018-ADG-835105 YoctoLHC. The computing resources have been brought to us by the Finnish IT Center for Science (CSC), under the project jyy2580. 

\bibliographystyle{JHEP}
\bibliography{SciPost_Example_BiBTeX_File.bib}

\providecommand{\href}[2]{#2}\begingroup\raggedright\begin{thebibliography}{10}

\bibitem{Combridge:1978kx}
B.~L. Combridge, {\it {Associated Production of Heavy Flavor States in p p and
  anti-p p Interactions: Some QCD Estimates}},  {\em Nucl. Phys. B} {\bf 151}
  (1979) 429--456.

\bibitem{Beenakker:1988bq}
W.~Beenakker, H.~Kuijf, W.~L. van Neerven, and J.~Smith, {\it {QCD Corrections
  to Heavy Quark Production in p anti-p Collisions}},  {\em Phys. Rev. D} {\bf
  40} (1989) 54--82.

\bibitem{Nason:1989zy}
P.~Nason, S.~Dawson, and R.~K. Ellis, {\it {The One Particle Inclusive
  Differential Cross-Section for Heavy Quark Production in Hadronic
  Collisions}},  {\em Nucl. Phys. B} {\bf 327} (1989) 49--92. [Erratum:
  Nucl.Phys.B 335, 260--260 (1990)].

\bibitem{Czakon:2013goa}
M.~Czakon, P.~Fiedler, and A.~Mitov, {\it {Total Top-Quark Pair-Production
  Cross Section at Hadron Colliders Through $O(\alpha^4_S)$}},  {\em Phys. Rev.
  Lett.} {\bf 110} (2013) 252004,
  [\href{http://xxx.lanl.gov/abs/1303.6254}{{\tt arXiv:1303.6254}}].

\bibitem{Catani:2020kkl}
S.~Catani, S.~Devoto, M.~Grazzini, S.~Kallweit, and J.~Mazzitelli, {\it
  {Bottom-quark production at hadron colliders: fully differential predictions
  in NNLO QCD}},  {\em JHEP} {\bf 03} (2021) 029,
  [\href{http://xxx.lanl.gov/abs/2010.1190}{{\tt arXiv:2010.1190}}].

\bibitem{Brodsky:1980pb}
S.~J. Brodsky, P.~Hoyer, C.~Peterson, and N.~Sakai, {\it {The Intrinsic Charm
  of the Proton}},  {\em Phys. Lett. B} {\bf 93} (1980) 451--455.

\bibitem{Ball:2022qks}
R.~D. Ball, A.~Candido, J.~Cruz-Martinez, S.~Forte, T.~Giani, F.~Hekhorn,
  K.~Kudashkin, G.~Magni, and J.~Rojo, {\it {Evidence for intrinsic charm
  quarks in the proton}},  {\em Nature} {\bf 608} (2022) 483--487,
  [\href{http://xxx.lanl.gov/abs/2208.0837}{{\tt arXiv:2208.0837}}].

\bibitem{Guzzi:2022rca}
M.~Guzzi, T.~J. Hobbs, K.~Xie, J.~Huston, P.~Nadolsky, and C.~P. Yuan, {\it
  {The persistent nonperturbative charm enigma}},
  \href{http://xxx.lanl.gov/abs/2211.0138}{{\tt arXiv:2211.0138}}.

\bibitem{Lyonnet:2015dca}
F.~Lyonnet, A.~Kusina, T.~Je\v{z}o, K.~Kovar\'\i{}k, F.~Olness, I.~Schienbein,
  and J.-Y. Yu, {\it {On the intrinsic bottom content of the nucleon and its
  impact on heavy new physics at the LHC}},  {\em JHEP} {\bf 07} (2015) 141,
  [\href{http://xxx.lanl.gov/abs/1504.0515}{{\tt arXiv:1504.0515}}].

\bibitem{Forte:2019hjc}
S.~Forte, T.~Giani, and D.~Napoletano, {\it {Fitting the b-quark PDF as a
  massive-b scheme: Higgs production in bottom fusion}},  {\em Eur. Phys. J. C}
  {\bf 79} (2019), no.~7 609, [\href{http://xxx.lanl.gov/abs/1905.0220}{{\tt
  arXiv:1905.0220}}].

\bibitem{Cacciari:2015fta}
M.~Cacciari, M.~L. Mangano, and P.~Nason, {\it {Gluon PDF constraints from the
  ratio of forward heavy-quark production at the LHC at $\sqrt{S}=7$ and 13
  TeV}},  {\em Eur. Phys. J. C} {\bf 75} (2015), no.~12 610,
  [\href{http://xxx.lanl.gov/abs/1507.0619}{{\tt arXiv:1507.0619}}].

\bibitem{Gauld:2015yia}
R.~Gauld, J.~Rojo, L.~Rottoli, and J.~Talbert, {\it {Charm production in the
  forward region: constraints on the small-x gluon and backgrounds for neutrino
  astronomy}},  {\em JHEP} {\bf 11} (2015) 009,
  [\href{http://xxx.lanl.gov/abs/1506.0802}{{\tt arXiv:1506.0802}}].

\bibitem{Eskola:2019bgf}
K.~J. Eskola, I.~Helenius, P.~Paakkinen, and H.~Paukkunen, {\it {A QCD analysis
  of LHCb D-meson data in p+Pb collisions}},  {\em JHEP} {\bf 05} (2020) 037,
  [\href{http://xxx.lanl.gov/abs/1906.0251}{{\tt arXiv:1906.0251}}].

\bibitem{Kusina:2020dki}
A.~Kusina, J.-P. Lansberg, I.~Schienbein, and H.-S. Shao, {\it {Reweighted
  nuclear PDFs using heavy-flavor production data at the LHC}},  {\em Phys.
  Rev. D} {\bf 104} (2021), no.~1 014010,
  [\href{http://xxx.lanl.gov/abs/2012.1146}{{\tt arXiv:2012.1146}}].

\bibitem{Bailey:2020ooq}
S.~Bailey, T.~Cridge, L.~A. Harland-Lang, A.~D. Martin, and R.~S. Thorne, {\it
  {Parton distributions from LHC, HERA, Tevatron and fixed target data: MSHT20
  PDFs}},  {\em Eur. Phys. J. C} {\bf 81} (2021), no.~4 341,
  [\href{http://xxx.lanl.gov/abs/2012.0468}{{\tt arXiv:2012.0468}}].

\bibitem{NNPDF:2021njg}
{\bf NNPDF} Collaboration, R.~D. Ball et~al., {\it {The path to proton
  structure at 1\% accuracy}},  {\em Eur. Phys. J. C} {\bf 82} (2022), no.~5
  428, [\href{http://xxx.lanl.gov/abs/2109.0265}{{\tt arXiv:2109.0265}}].

\bibitem{Hou:2019efy}
T.-J. Hou et~al., {\it {New CTEQ global analysis of quantum chromodynamics with
  high-precision data from the LHC}},  {\em Phys. Rev. D} {\bf 103} (2021),
  no.~1 014013, [\href{http://xxx.lanl.gov/abs/1912.1005}{{\tt
  arXiv:1912.1005}}].

\bibitem{CMS:2016jip}
{\bf CMS} Collaboration, V.~Khachatryan et~al., {\it {Measurement of the
  double-differential inclusive jet cross section in proton\textendash{}proton
  collisions at $\sqrt{s} = 13\,\text {TeV} $}},  {\em Eur. Phys. J. C} {\bf
  76} (2016), no.~8 451, [\href{http://xxx.lanl.gov/abs/1605.0443}{{\tt
  arXiv:1605.0443}}].

\bibitem{ALICE:2019qyj}
{\bf ALICE} Collaboration, S.~Acharya et~al., {\it {Measurements of inclusive
  jet spectra in pp and central Pb-Pb collisions at $\sqrt{s_{\rm{NN}}}$ = 5.02
  TeV}},  {\em Phys. Rev. C} {\bf 101} (2020), no.~3 034911,
  [\href{http://xxx.lanl.gov/abs/1909.0971}{{\tt arXiv:1909.0971}}].

\bibitem{ALICE:2022cxs}
{\bf ALICE} Collaboration, {\it {W$^\pm$-boson production in p$-$Pb collisions
  at $\sqrt{s_{NN}} = 8.16$ TeV and PbPb collisions at $\sqrt{s_{NN}} = 5.02$
  TeV}},  \href{http://xxx.lanl.gov/abs/2204.1064}{{\tt arXiv:2204.1064}}.

\bibitem{CMS:2017uoy}
{\bf CMS} Collaboration, A.~M. Sirunyan et~al., {\it {Measurement of the
  ${B}^{\pm}$ Meson Nuclear Modification Factor in Pb-Pb Collisions at
  $\sqrt{{s}_{NN}}=5.02\text{ }\text{ }\mathrm{TeV}$}},  {\em Phys. Rev. Lett.}
  {\bf 119} (2017), no.~15 152301,
  [\href{http://xxx.lanl.gov/abs/1705.0472}{{\tt arXiv:1705.0472}}].

\bibitem{CDF:2004jtw}
{\bf CDF} Collaboration, D.~Acosta et~al., {\it {Measurement of the $J/\psi$
  meson and $b-$hadron production cross sections in $p\bar{p}$ collisions at
  $\sqrt{s} = 1960$ GeV}},  {\em Phys. Rev. D} {\bf 71} (2005) 032001,
  [\href{http://xxx.lanl.gov/abs/hep-ex/0412071}{{\tt hep-ex/0412071}}].

\bibitem{CDF:2006ipg}
{\bf CDF} Collaboration, A.~Abulencia et~al., {\it {Measurement of the B+
  production cross-section in p anti-p collisions at s**(1/2) = 1960-GeV}},
  {\em Phys. Rev. D} {\bf 75} (2007) 012010,
  [\href{http://xxx.lanl.gov/abs/hep-ex/0612015}{{\tt hep-ex/0612015}}].

\bibitem{CMS:2011pdu}
{\bf CMS} Collaboration, S.~Chatrchyan et~al., {\it {Measurement of the $B^0$
  production cross section in $pp$ Collisions at $\sqrt{s}=7$ TeV}},  {\em
  Phys. Rev. Lett.} {\bf 106} (2011) 252001,
  [\href{http://xxx.lanl.gov/abs/1104.2892}{{\tt arXiv:1104.2892}}].

\bibitem{CMS:2011oft}
{\bf CMS} Collaboration, V.~Khachatryan et~al., {\it {Measurement of the $B^+$
  Production Cross Section in pp Collisions at $\sqrt{s} =
  7$\textasciitilde{}TeV}},  {\em Phys. Rev. Lett.} {\bf 106} (2011) 112001,
  [\href{http://xxx.lanl.gov/abs/1101.0131}{{\tt arXiv:1101.0131}}].

\bibitem{CMS:2011kew}
{\bf CMS} Collaboration, S.~Chatrchyan et~al., {\it {Measurement of the Strange
  $B$ Meson Production Cross Section with J/Psi $\phi$ Decays in $pp$
  Collisions at $\sqrt{s}=7$ TeV}},  {\em Phys. Rev. D} {\bf 84} (2011) 052008,
  [\href{http://xxx.lanl.gov/abs/1106.4048}{{\tt arXiv:1106.4048}}].

\bibitem{LHCb:2012sng}
{\bf LHCb} Collaboration, R.~Aaij et~al., {\it {Measurement of the $B^\pm$
  production cross-section in $pp$ collisions at $\sqrt{s}=7$ TeV}},  {\em
  JHEP} {\bf 04} (2012) 093, [\href{http://xxx.lanl.gov/abs/1202.4812}{{\tt
  arXiv:1202.4812}}].

\bibitem{LHCb:2013vjr}
{\bf LHCb} Collaboration, R.~Aaij et~al., {\it {Measurement of B meson
  production cross-sections in proton-proton collisions at $\sqrt{s}$ = 7
  TeV}},  {\em JHEP} {\bf 08} (2013) 117,
  [\href{http://xxx.lanl.gov/abs/1306.3663}{{\tt arXiv:1306.3663}}].

\bibitem{ATLAS:2013cia}
{\bf ATLAS} Collaboration, G.~Aad et~al., {\it {Measurement of the differential
  cross-section of $B^{+}$ meson production in pp collisions at $\sqrt{s}$ = 7
  TeV at ATLAS}},  {\em JHEP} {\bf 10} (2013) 042,
  [\href{http://xxx.lanl.gov/abs/1307.0126}{{\tt arXiv:1307.0126}}].

\bibitem{CMS:2016plw}
{\bf CMS} Collaboration, V.~Khachatryan et~al., {\it {Measurement of the total
  and differential inclusive $B^+$ hadron cross sections in pp collisions at
  $\sqrt{s}$ = 13 TeV}},  {\em Phys. Lett. B} {\bf 771} (2017) 435--456,
  [\href{http://xxx.lanl.gov/abs/1609.0087}{{\tt arXiv:1609.0087}}].

\bibitem{LHCb:2016qpe}
{\bf LHCb} Collaboration, R.~Aaij et~al., {\it {Measurement of the $b$-quark
  production cross-section in 7 and 13 TeV $pp$ collisions}},  {\em Phys. Rev.
  Lett.} {\bf 118} (2017), no.~5 052002,
  [\href{http://xxx.lanl.gov/abs/1612.0514}{{\tt arXiv:1612.0514}}]. [Erratum:
  Phys.Rev.Lett. 119, 169901 (2017)].

\bibitem{LHCb:2017vec}
{\bf LHCb} Collaboration, R.~Aaij et~al., {\it {Measurement of the $B^{\pm}$
  production cross-section in pp collisions at $\sqrt{s} =$ 7 and 13 TeV}},
  {\em JHEP} {\bf 12} (2017) 026,
  [\href{http://xxx.lanl.gov/abs/1710.0492}{{\tt arXiv:1710.0492}}].

\bibitem{CMS:2015sfx}
{\bf CMS} Collaboration, V.~Khachatryan et~al., {\it {Study of B Meson
  Production in p$+$Pb Collisions at $\sqrt{s_{NN}}=5.02$ TeV Using Exclusive
  Hadronic Decays}},  {\em Phys. Rev. Lett.} {\bf 116} (2016), no.~3 032301,
  [\href{http://xxx.lanl.gov/abs/1508.0667}{{\tt arXiv:1508.0667}}].

\bibitem{LHCb:2019avm}
{\bf LHCb} Collaboration, R.~Aaij et~al., {\it {Measurement of $B^+$, $B^0$ and
  $\Lambda_b^0$ production in $p\mkern 1mu\mathrm{Pb}$ collisions at
  $\sqrt{s_\mathrm{NN}}=8.16\,{\rm TeV}$}},  {\em Phys. Rev. D} {\bf 99}
  (2019), no.~5 052011, [\href{http://xxx.lanl.gov/abs/1902.0559}{{\tt
  arXiv:1902.0559}}].

\bibitem{Thorne:2008xf}
R.~S. Thorne and W.~K. Tung, {\it {PQCD Formulations with Heavy Quark Masses
  and Global Analysis}},  in {\em {HERA and the LHC: 4th Workshop on the
  Implications of HERA for LHC Physics}}, pp.~332--351, 9, 2008.
\newblock \href{http://xxx.lanl.gov/abs/0809.0714}{{\tt arXiv:0809.0714}}.

\bibitem{Cacciari:1998it}
M.~Cacciari, M.~Greco, and P.~Nason, {\it {The P(T) spectrum in heavy flavor
  hadroproduction}},  {\em JHEP} {\bf 05} (1998) 007,
  [\href{http://xxx.lanl.gov/abs/hep-ph/9803400}{{\tt hep-ph/9803400}}].

\bibitem{Kniehl:2004fy}
B.~A. Kniehl, G.~Kramer, I.~Schienbein, and H.~Spiesberger, {\it {Inclusive
  D*+- production in p anti-p collisions with massive charm quarks}},  {\em
  Phys. Rev. D} {\bf 71} (2005) 014018,
  [\href{http://xxx.lanl.gov/abs/hep-ph/0410289}{{\tt hep-ph/0410289}}].

\bibitem{Kniehl:2005mk}
B.~A. Kniehl, G.~Kramer, I.~Schienbein, and H.~Spiesberger, {\it {Collinear
  subtractions in hadroproduction of heavy quarks}},  {\em Eur. Phys. J. C}
  {\bf 41} (2005) 199--212, [\href{http://xxx.lanl.gov/abs/hep-ph/0502194}{{\tt
  hep-ph/0502194}}].

\bibitem{Kniehl:2007erq}
B.~A. Kniehl, G.~Kramer, I.~Schienbein, and H.~Spiesberger, {\it {Finite-mass
  effects on inclusive $B$ meson hadroproduction}},  {\em Phys. Rev. D} {\bf
  77} (2008) 014011, [\href{http://xxx.lanl.gov/abs/0705.4392}{{\tt
  arXiv:0705.4392}}].

\bibitem{Kniehl:2011bk}
B.~A. Kniehl, G.~Kramer, I.~Schienbein, and H.~Spiesberger, {\it {Inclusive
  B-Meson Production at the LHC in the GM-VFN Scheme}},  {\em Phys. Rev. D}
  {\bf 84} (2011) 094026, [\href{http://xxx.lanl.gov/abs/1109.2472}{{\tt
  arXiv:1109.2472}}].

\bibitem{Kniehl:2015fla}
B.~A. Kniehl, G.~Kramer, I.~Schienbein, and H.~Spiesberger, {\it {Inclusive
  $B$-meson production at small $p_T$ in the general-mass
  variable-flavor-number scheme}},  {\em Eur. Phys. J. C} {\bf 75} (2015),
  no.~3 140, [\href{http://xxx.lanl.gov/abs/1502.0100}{{\tt arXiv:1502.0100}}].

\bibitem{Kramer:2018vde}
G.~Kramer and H.~Spiesberger, {\it {$b$-hadron production in the general-mass
  variable-flavour-number scheme and LHC data}},  {\em Phys. Rev. D} {\bf 98}
  (2018), no.~11 114010, [\href{http://xxx.lanl.gov/abs/1809.0429}{{\tt
  arXiv:1809.0429}}].

\bibitem{Benzke:2019usl}
M.~Benzke, B.~A. Kniehl, G.~Kramer, I.~Schienbein, and H.~Spiesberger, {\it
  {B-meson production in the general-mass variable-flavour-number scheme and
  LHC data}},  {\em Eur. Phys. J. C} {\bf 79} (2019), no.~10 814,
  [\href{http://xxx.lanl.gov/abs/1907.1245}{{\tt arXiv:1907.1245}}].

\bibitem{Helenius:2018uul}
I.~Helenius and H.~Paukkunen, {\it {Revisiting the D-meson hadroproduction in
  general-mass variable flavour number scheme}},  {\em JHEP} {\bf 05} (2018)
  196, [\href{http://xxx.lanl.gov/abs/1804.0355}{{\tt arXiv:1804.0355}}].

\bibitem{Guzzi:2011ew}
M.~Guzzi, P.~M. Nadolsky, H.-L. Lai, and C.~P. Yuan, {\it {General-Mass
  Treatment for Deep Inelastic Scattering at Two-Loop Accuracy}},  {\em Phys.
  Rev. D} {\bf 86} (2012) 053005,
  [\href{http://xxx.lanl.gov/abs/1108.5112}{{\tt arXiv:1108.5112}}].

\bibitem{Xie:2019eoe}
K.~Xie, {\em {Massive elementary particles in the Standard Model and its
  supersymmetric triplet Higgs extension}}.
\newblock PhD thesis, Southern Methodist U., 2019.

\bibitem{Xie:2021ycd}
K.~Xie, J.~M. Campbell, and P.~M. Nadolsky, {\it {A general-mass scheme for
  prompt charm production at hadron colliders}},  {\em SciPost Phys. Proc.}
  {\bf 8} (2022) 084, [\href{http://xxx.lanl.gov/abs/2108.0374}{{\tt
  arXiv:2108.0374}}].

\bibitem{Nason:2004rx}
P.~Nason, {\it {A New method for combining NLO QCD with shower Monte Carlo
  algorithms}},  {\em JHEP} {\bf 11} (2004) 040,
  [\href{http://xxx.lanl.gov/abs/hep-ph/0409146}{{\tt hep-ph/0409146}}].

\bibitem{Frixione:2007vw}
S.~Frixione, P.~Nason, and C.~Oleari, {\it {Matching NLO QCD computations with
  Parton Shower simulations: the POWHEG method}},  {\em JHEP} {\bf 11} (2007)
  070, [\href{http://xxx.lanl.gov/abs/0709.2092}{{\tt arXiv:0709.2092}}].

\bibitem{Alioli:2010xd}
S.~Alioli, P.~Nason, C.~Oleari, and E.~Re, {\it {A general framework for
  implementing NLO calculations in shower Monte Carlo programs: the POWHEG
  BOX}},  {\em JHEP} {\bf 06} (2010) 043,
  [\href{http://xxx.lanl.gov/abs/1002.2581}{{\tt arXiv:1002.2581}}].

\bibitem{Mazzitelli:2023znt}
J.~Mazzitelli, A.~Ratti, M.~Wiesemann, and G.~Zanderighi, {\it {B-hadron
  production at the LHC from bottom-quark pair production at NNLO+PS}},
  \href{http://xxx.lanl.gov/abs/2302.0164}{{\tt arXiv:2302.0164}}.

\bibitem{Albino:2008fy}
S.~Albino, B.~A. Kniehl, and G.~Kramer, {\it {AKK Update: Improvements from New
  Theoretical Input and Experimental Data}},  {\em Nucl. Phys. B} {\bf 803}
  (2008) 42--104, [\href{http://xxx.lanl.gov/abs/0803.2768}{{\tt
  arXiv:0803.2768}}].

\bibitem{Aversa:1988vb}
F.~Aversa, P.~Chiappetta, M.~Greco, and J.~P. Guillet, {\it {QCD Corrections to
  Parton-Parton Scattering Processes}},  {\em Nucl. Phys. B} {\bf 327} (1989)
  105.

\bibitem{Mele:1990cw}
B.~Mele and P.~Nason, {\it {The Fragmentation function for heavy quarks in
  QCD}},  {\em Nucl. Phys. B} {\bf 361} (1991) 626--644. [Erratum: Nucl.Phys.B
  921, 841--842 (2017)].

\bibitem{Melnikov:2004bm}
K.~Melnikov and A.~Mitov, {\it {Perturbative heavy quark fragmentation function
  through $\mathcal{O}(\alpha^2_s)$}},  {\em Phys. Rev. D} {\bf 70} (2004)
  034027, [\href{http://xxx.lanl.gov/abs/hep-ph/0404143}{{\tt
  hep-ph/0404143}}].

\bibitem{Collins:1978wz}
J.~C. Collins, F.~Wilczek, and A.~Zee, {\it {Low-Energy Manifestations of Heavy
  Particles: Application to the Neutral Current}},  {\em Phys. Rev. D} {\bf 18}
  (1978) 242.

\bibitem{SLD:2002poq}
{\bf SLD} Collaboration, K.~Abe et~al., {\it {Measurement of the b quark
  fragmentation function in Z0 decays}},  {\em Phys. Rev. D} {\bf 65} (2002)
  092006, [\href{http://xxx.lanl.gov/abs/hep-ex/0202031}{{\tt
  hep-ex/0202031}}]. [Erratum: Phys.Rev.D 66, 079905 (2002)].

\bibitem{OPAL:2002plk}
{\bf OPAL} Collaboration, G.~Abbiendi et~al., {\it {Inclusive analysis of the b
  quark fragmentation function in Z decays at LEP}},  {\em Eur. Phys. J. C}
  {\bf 29} (2003) 463--478, [\href{http://xxx.lanl.gov/abs/hep-ex/0210031}{{\tt
  hep-ex/0210031}}].

\bibitem{ALEPH:2001pfo}
{\bf ALEPH} Collaboration, A.~Heister et~al., {\it {Study of the fragmentation
  of b quarks into B mesons at the Z peak}},  {\em Phys. Lett. B} {\bf 512}
  (2001) 30--48, [\href{http://xxx.lanl.gov/abs/hep-ex/0106051}{{\tt
  hep-ex/0106051}}].

\bibitem{Czakon:2022pyz}
M.~Czakon, T.~Generet, A.~Mitov, and R.~Poncelet, {\it {NNLO B-fragmentation
  fits and their application to $t\bar t$ production and decay at the LHC}},
  \href{http://xxx.lanl.gov/abs/2210.0607}{{\tt arXiv:2210.0607}}.

\bibitem{Cridge:2021qfd}
T.~Cridge, L.~A. Harland-Lang, A.~D. Martin, and R.~S. Thorne, {\it {An
  investigation of the $\alpha _S$ and heavy quark mass dependence in the
  MSHT20 global PDF analysis}},  {\em Eur. Phys. J. C} {\bf 81} (2021), no.~8
  744, [\href{http://xxx.lanl.gov/abs/2106.1028}{{\tt arXiv:2106.1028}}].

\bibitem{Eskola:2021nhw}
K.~J. Eskola, P.~Paakkinen, H.~Paukkunen, and C.~A. Salgado, {\it {EPPS21: A
  global QCD analysis of nuclear PDFs}},
  \href{http://xxx.lanl.gov/abs/2112.1246}{{\tt arXiv:2112.1246}}.

\bibitem{Khalek:2022zqe}
R.~A. Khalek, R.~Gauld, T.~Giani, E.~R. Nocera, T.~R. Rabemananjara, and
  J.~Rojo, {\it {nNNPDF3.0: Evidence for a modified partonic structure in heavy
  nuclei}},  \href{http://xxx.lanl.gov/abs/2201.1236}{{\tt arXiv:2201.1236}}.

\bibitem{LHCb:2015swx}
{\bf LHCb} Collaboration, R.~Aaij et~al., {\it {Measurements of prompt charm
  production cross-sections in $pp$ collisions at $ \sqrt{s}=13 $ TeV}},  {\em
  JHEP} {\bf 03} (2016) 159, [\href{http://xxx.lanl.gov/abs/1510.0170}{{\tt
  arXiv:1510.0170}}]. [Erratum: JHEP 09, 013 (2016), Erratum: JHEP 05, 074
  (2017)].

\bibitem{Cacciari:2012ny}
M.~Cacciari, S.~Frixione, N.~Houdeau, M.~L. Mangano, P.~Nason, and G.~Ridolfi,
  {\it {Theoretical predictions for charm and bottom production at the LHC}},
  {\em JHEP} {\bf 10} (2012) 137,
  [\href{http://xxx.lanl.gov/abs/1205.6344}{{\tt arXiv:1205.6344}}].

\bibitem{FONLL:web}
``The fonll web interface.''
  \url{http://www.lpthe.jussieu.fr/~cacciari/fonll/fonllform.html}.
\newblock Accessed: Jan. 2023.

\bibitem{NNPDF:2014otw}
{\bf NNPDF} Collaboration, R.~D. Ball et~al., {\it {Parton distributions for
  the LHC Run II}},  {\em JHEP} {\bf 04} (2015) 040,
  [\href{http://xxx.lanl.gov/abs/1410.8849}{{\tt arXiv:1410.8849}}].

\bibitem{LHCb:2017yua}
{\bf LHCb} Collaboration, R.~Aaij et~al., {\it {Study of prompt D$^{0}$ meson
  production in $p$Pb collisions at $ \sqrt{s_{\mathrm{NN}}}=5 $ TeV}},  {\em
  JHEP} {\bf 10} (2017) 090, [\href{http://xxx.lanl.gov/abs/1707.0275}{{\tt
  arXiv:1707.0275}}].

\bibitem{Bierlich:2022pfr}
C.~Bierlich et~al., {\it {A comprehensive guide to the physics and usage of
  PYTHIA 8.3}},  {\em SciPost Phys. Codebases} {\bf 08} (3, 2022)
  [\href{http://xxx.lanl.gov/abs/2203.1160}{{\tt arXiv:2203.1160}}].

\bibitem{Duwentaster:2022kpv}
P.~Duwent\"aster, T.~Je\v{z}o, M.~Klasen, K.~Kova\v{r}\'\i{}k, A.~Kusina, K.~F.
  Muzakka, F.~I. Olness, R.~Ruiz, I.~Schienbein, and J.~Y. Yu, {\it {Impact of
  heavy quark and quarkonium data on nuclear gluon PDFs}},  {\em Phys. Rev. D}
  {\bf 105} (2022), no.~11 114043,
  [\href{http://xxx.lanl.gov/abs/2204.0998}{{\tt arXiv:2204.0998}}].

\bibitem{Courtoy:2022ocu}
A.~Courtoy, J.~Huston, P.~Nadolsky, K.~Xie, M.~Yan, and C.~P. Yuan, {\it
  {Parton distributions need representative sampling}},
  \href{http://xxx.lanl.gov/abs/2205.1044}{{\tt arXiv:2205.1044}}.

\end{thebibliography}\endgroup


\end{document}